\documentclass[aps,prb,reprint,twocolumn,showpacs,superscriptaddress]{revtex4-2}
\usepackage{graphicx}
\usepackage{amsmath}
\usepackage{amsthm}
\usepackage{bm}
\usepackage{float}
\usepackage{txfonts}
\usepackage{amsfonts}
\usepackage{amssymb}
\usepackage{array}
\usepackage{multirow,bigdelim}
\graphicspath{{./figures/}}

\usepackage[colorlinks=true,citecolor=blue,urlcolor=blue]{hyperref}

\begin{document}

\title{Digital quantum simulation of the Su-Schrieffer-Heeger model 
using a parameterized quantum circuit}

\author{Qing~Xie}
\affiliation{Quantum Computational Science Research Team, RIKEN Center for Quantum Computing (RQC), Saitama 351-0198, Japan}

\author{Kazuhiro~Seki}
\affiliation{Quantum Computational Science Research Team, RIKEN Center for Quantum Computing (RQC), Saitama 351-0198, Japan}

\author{Tomonori~Shirakawa}
\affiliation{Quantum Computational Science Research Team, RIKEN Center for Quantum Computing (RQC), Saitama 351-0198, Japan}
\affiliation{Computational Materials Science Research Team, RIKEN Center for Computational Science (R-CCS), Hyogo 650-0047, Japan}
\affiliation{Computational Condensed Matter Physics Laboratory, RIKEN Cluster for Pioneering Research (CPR), Saitama 351-0198, Japan}
\affiliation{RIKEN Interdisciplinary Theoretical and Mathematical Sciences Program (iTHEMS), Wako, Saitama 351-0198, Japan}

\author{Seiji~Yunoki}
\affiliation{Quantum Computational Science Research Team, RIKEN Center for Quantum Computing (RQC), Saitama 351-0198, Japan}
\affiliation{Computational Materials Science Research Team, RIKEN Center for Computational Science (R-CCS),  Hyogo 650-0047,  Japan}
\affiliation{Computational Condensed Matter Physics Laboratory, RIKEN Cluster for Pioneering Research (CPR), Saitama 351-0198, Japan}
\affiliation{Computational Quantum Matter Research Team, RIKEN Center for Emergent Matter Science (CEMS), Saitama 351-0198, Japan}


\date{\today}

\begin{abstract}

We perform digital quantum simulations of the noninteracting Su–Schrieffer–Heeger (SSH) model using a parameterized quantum circuit. 
The circuit comprises two main components: the first prepares the initial state from the product state $|0\rangle^{\otimes L}$, where $L$ is the system size; the second consists of $M$ layers of brick-wall unitaries simulating time evolution. 
The evolution times, encoded as the rotation angles of quantum gates in the second part, are optimized variationally to minimize the energy. 
The SSH model exhibits two distinct topological phases, depending on the relative strengths of inter- and intra-cell hopping amplitudes. 
We investigate the evolution of the energy, entanglement entropy, and mutual information towards topologically trivial and nontrivial ground states. 
Our results find the follows: (i) When the initial and target ground states belong to the same topological phase, the variational energy decreases exponentially, the entanglement entropy quickly saturates in a system-size-independent manner, and the mutual information remains spatially localized, as the number of layers increases. 
(ii) When the initial and target ground states belong to different topological phases, the variational energy decreases polynomially, the entanglement entropy initially grows logarithmically before decreasing, and the mutual information spreads ballistically across the entire system, with increasing the number of layers.
Furthermore, by calculating the polarization, we identify a topological phase transition occurring at an intermediate circuit layer when the initial and final target states lie in different topological characters. Finally, we experimentally confirm this topological phase transition in an 18-site system using 19 qubits on a trapped-ion quantum computer provided by Quantinuum.

\end{abstract}

\maketitle



\section{Introduction}

Simulating quantum many-body systems on classical hardware is 
notoriously challenging due to the exponential growth of the Hilbert 
space, rendering exact solutions computationally intractable for 
large systems. This intrinsic complexity highlights a compelling use 
for quantum computers and quantum algorithms, which exploit 
quantum mechanical principles to simulate quantum dynamics more 
efficiently than their classical 
counterparts~\cite{Feynman1982, Lloyd96}. 
In recent years, there has been growing interest in 
both analog and digital quantum simulations of quantum many-body 
systems using various quantum hardware platforms. These simulations 
aim to explore fundamental problems in physics, such as 
strongly correlated quantum systems~\cite{Georgescu14, Mazurenko2017, Seki2020, Stanisic_2022, Sun2023, Kikuchi2023, Maskara2023, Summer2024, Hemery2024, Seki2024},
lattice gauge theories~\cite{Banuls20, Farrell2024, Hayata2024, Hayata2025}, 
and topological phases of matter~\cite{Zhang2018, Satzinger2021, Smith2022}. 
The rapid advancement of noisy intermediate-scale quantum (NISQ) 
devices~\cite{Moses2023, McKay2023, Bluvstein2023, Sunada2024, Li2024, Gao2024}, 
as introduced by Preskill~\cite{Preskill2018}, has enabled several 
proof-of-principle demonstrations of quantum supremacy or advantage~\cite{Arute2019, Zhong20, Wu21,
Zhong21, Morvan2024, DeCross2024}, paving the way for 
simulating complex quantum phenomena beyond the capability of 
classical computers~\cite{Kim2023, Alexeev2024,
Shinjo24, Moreno24}.

The development of quantum-classical hybrid algorithms has been 
instrumental in advancing quantum computation in the NISQ era. 
In particular, variational quantum algorithms (VQAs), 
such as the variational quantum eigensolver~\cite{Peruzzo_2014, Kandala_2017} 
and the quantum approximate optimization algorithm~\cite{Farhi14}, 
have found widespread applications in physics and chemistry. 
A central challenge in VQAs is the design of an effective variational 
ansatz, which critically affects both the efficiency and accuracy 
of quantum computation. For example, 
ansatze that respect (some of) the symmetries of the 
Hamiltonian~\cite{Seki2020, Seki2022_PRA} or topological 
sectors~\cite{Sun2023, Sun_2023_loopgas} can be lead to more 
efficient ground-state preparation of quantum many-body systems.

In addition to the selection of ansatz, studying its fundamental 
properties is equally important. One prominent challenge is 
the phenomenon of barren plateaus~\cite{mcclean_barren_2018}, 
where gradients in high-dimensional parameter spaces vanish 
exponentially with the circuit depth or system size, making 
optimization exponentially difficult. 
A thorough investigation of the discretized quantum adiabatic process 
(DQAP) ansatz for noninteracting fermions in one dimension has
demonstrated that an appropriate choice of classical optimization 
scheme can alleviate these difficulties in this particular case, 
leading to systematic convergence to the optimal variational 
parameters~\cite{PhysRevResearch.3.013004}. Similarly, an detailed 
study of a variational ansatz for an effective spin-1 chain has 
revealed a relationship between the ansatz's expressibility 
and the spin correlation 
length in the symmetry-protected topological Haldane phase, showing 
that the accuracy of the ansatz is governed not by the system size, 
but by the correlation length~\cite{Sun2023}.

This work investigates the fundamental properties of the
DQAP~ansatz~\cite{PhysRevResearch.3.013004} for preparing 
ground states of the Su–Schrieffer–Heeger (SSH) 
model~\cite{PhysRevLett.42.1698} at half filling, a
paradigmatic system in the study of topological phase 
transitions~\cite{Asboth2015}.
Starting from a topologically trivial product state, the ansatz 
generates target states in either the topologically trivial or 
nontrivial phase, depending on the model parameters.
Our numerical simulations show that topological characters of 
the initial and final target states lead to qualitative differences 
in the evolution of the energy, entanglement entropy,
and mutual information, as summarized in Table~\ref{tab}. 
The topological nature of the DQAP ansatz is further characterized 
by calculating the polarization as a function of circuit depth. 
Our results reveal that a topological phase transition occurs at 
a critical circuit depth, even before reaching the target ground 
state. 
Finally, we validate our numerical findings using the 20-qubit
Quantinuum H1-1 trapped-ion device~\cite{H1datasheet} to simulate 
an 18-site SSH model. The high fidelity of long-range two-qubit 
gates, enabled by the all-to-all connectivity architecture of the 
H1-1 system, makes it particularly well-suited for this purpose.

\begin{table*}
\caption{Summary of the behavior of four quantities--energy, entanglement entropy, mutual
information, and polarization--during the DQAP evolution of the optimized DQAP ansatz
$|\Psi_M(\bm\theta)\rangle$, depending on whether the initial and final states belong to
the same or different topological phases. The corresponding figures are referenced in the
last column.
Note that the number of layers required to exactly represent the ground state of the
final Hamiltonian is $L/4$ [$(L-2)/4$], irrespective of the topological nature of the
initial and final states, assuming that APBCs (PBCs) are applied with $L\in 4\mathbb{N}$
($L\in4\mathbb{N}+2$), and $M^*$ is less than this number of layers. \label{tab}}
\begin{tabular}{l|lll}
\hline \hline
Initial and final states &    Same phase   &  Different phases  & Fig.~\ref{fig:fig1} \\
\hline
Energy & exponential decrease in $M$  & polynomial decrease in $M$ &   Fig.~\ref{fig:fig3} \\
Entanglement entropy & monotonic increase in $M$  & nonmonotonic dependence on $M$  &  Fig.~\ref{fig:fig4} \\
Mutual information   & spatially confined  &  spatially spread with $M$  & Fig.~\ref{fig:fig6}  \\
Polarization & ---  & alternation at $M^\ast$
& Figs.~\ref{fig:fig7}, \ref{fig:fig8}, \ref{fig:figReIm} \\
\hline \hline
\end{tabular}
\end{table*}

The remainder of this paper is organized as follows.
In Sec.~\ref{sec:model}, we define the SSH model and introduce the 
parameterized quantum circuit based on the DQAP ansatz.
Section~\ref{sec:results} presents numerical results on the 
variational ground-state energy, entanglement entropy, mutual 
information, and polarization, characterizing the 
evolution process from the initial to final state under the DQAP 
ansatz.
In Sec.~\ref{sec:qc}, we demonstrate how a quantum computer is used 
to distinguish topologically distinct phases by computing the 
polarization.
Finally, Sec.~\ref{sec:conclusion} summarizes our findings and 
provides concluding discussions. Additional numerical results, 
details of the quantum circuit implemented on hardware, and 
supplementary experimental results are provided in
Appendices~\ref{app:results}, \ref{app:native2Qgates}, 
and \ref{app:experiments}, respectively.

\section{Model and method} \label{sec:model}

\begin{figure}
\includegraphics[width=0.98\columnwidth]{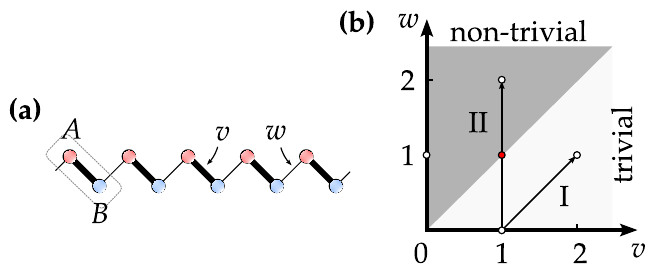}
\caption{
(a) A schematic representation of the 1D SSH model for spinless fermions. $A$
and $B$ denote two sublattices within a unit cell. $v$ and $w$ represent the intra-cell
and inter-cell hopping amplitudes, respectively. (b) The phase diagram at half-filling.
Two evolutionary paths are indicated by lines with arrows. Path I corresponds to a case
where both the initial and final states belong to the trivial phase, while Path II
represents a transition to a topologically nontrivial phase. The red dot marks the
critical point studied in Ref.~\cite{PhysRevResearch.3.013004}.
}
\label{fig:fig1}
\end{figure}

We consider the 1D SSH model on a ring with an even number $L$ of sites, as shown in
Fig.~\ref{fig:fig1}(a). The Hamiltonian is given by
\begin{align}
\hat{\mathcal{H}}_{\text{SSH}} = - \sum_{i=1}^{L/2}
\left( v \hat{c}^\dag_{A,i} \hat{c}^{}_{B,i} + \gamma w \hat{c}^\dag_{B,i} \hat{c}^{}_{A,i+1}  \right) + \text{H.c.},
\end{align}
where $A$ and $B$ denote the two sublattices within a unit cell, $\hat c_{A(B), i}^\dag$
represents the creation operator for a spinless fermion at the $i$th unit cell on
sublattice $A$ ($B$), and $v$ and $w$ are the intra-cell and inter-cell hopping
amplitudes, respectively. For convenience, we refer to them as $v$-bands and $w$-bonds.
The parameter $\gamma$ is introduced to account for different boundary conditions. For
hopping across the boundary, between the first and the last sites, we set $\gamma=1$ for
periodic boundary conditions (PBCs) and $\gamma=-1$ for anti-periodic boundary conditions
(APBCs). For all other hoppings that do not cross the boundary, we assume $\gamma=1$. In
this paper, we consider the system at half-filling, implying that the density of spinless
fermions is $L/2$. As shown in Fig.~\ref{fig:fig1}(b), it is well know~\cite{Asboth2015}
that the system exhibits a trivial insulating phase when $v/w > 1$. In contrast, for
$v/w < 1$, the ground state is in a topological phase characterized by a nontrivial
polarization~\cite{PhysRevLett.80.1800} under PBCs or APBCs. Under open-boundary
conditions, this phase supports two edge modes, which are protected by chiral
symmetry~\cite{PhysRevLett.42.1698}. The critical point at $v/w = 1$ has been previously
studied in Ref.~\cite{PhysRevResearch.3.013004} using a parameterized quantum circuit
approach, which we also employ in this work.

To simulate a fermionic Hamiltonian on a quantum circuit, the first step is to map it
onto a qubit basis. Several transformations can be employed for this purpose, including
the Jordan-Wigner transformation~\cite{ZeitschriftfurPhysik.47.631}, the Bravyi-Kitave
transformation~\cite{BRAVYI2002210}, the Ball-Verstracete-Cirac
transformation~\cite{PhysRevLett.95.176407, JStatMechTheoryExp.2005.P09012}, and other
recently developed methods~\cite{2022arXiv220909225R, NewJPhys.20.063010,
Jiang2020optimalfermionto, PRXQuantum.4.010326, Nys2023quantumcircuits}. In this work, we
adopt the Jordan-Wigner transformation (JWT)~\cite{PhysRevResearch.3.013004,
ZeitschriftfurPhysik.47.631}, which maps fermionic opertors to qubit operators as follows:
\begin{align}
\hat{c}_i^\dag & = \sigma_{i}^{-}\hat{K} , ~~~\hat{c}_i  = \hat{K} \sigma_{i}^{+},
\end{align}
where $\sigma^{\pm}_i = \frac{1}{2}(\hat{X}_i \pm i \hat{Y}_i)$ and
$\hat{K} = \exp[- i \frac{\pi}{2} \sum_{j < i}(\hat{Z}_j + 1)]$ is a nonlocal string
operator. Here, $\hat{X}_i, \hat{Y}_i, \hat{Z}_i$ denotes the Pauli matrices acting on
site $i$, with the sublattice index omitted for simplicity.

\begin{figure}
\includegraphics[width=0.95\columnwidth]{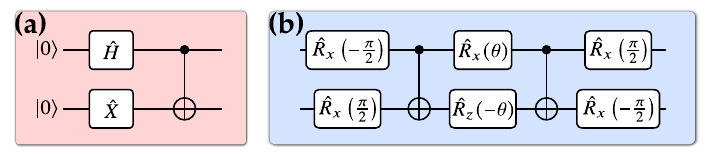}
\caption{
Quantum circuit implementations for (a) the initial state
 $|\mathrm{t}\rangle =\frac{1}{\sqrt{2}}(|01\rangle + |10\rangle)$ and (b) the unitary time
 evolution operator $\exp[-i\frac{\theta}{2}(\hat{X}_i\hat{X}_j +\hat{Y}_i\hat{Y}_j)]$. In
 the circuits, $\hat{X}$, $\hat{H}$ denote the Pauli-$X$ and Hadamard gates, respectively,
 while the rotation gates are defined as
 $\hat{R}_x(\theta) =\exp(-i\frac{\theta}{2}\hat{X})$ and
 $\hat{R}_z(\theta) =\exp(-i\frac{\theta}{2}\hat{Z})$. The~\textsc{Cnot} gates are
 represented in the standard form.
}
\label{fig:fig2}
\end{figure}

For the 1D SSH model, which contains only nearest-neighbor hopping terms, the JWT yields
the following mapping:
\begin{align}
\hat{c}^{\dag}_i \hat{c}^{}_{i+1} + \text{H.c.} \mapsto \sigma_{i}^{+} \sigma_{i+1}^{-} +   \text{H.c.}
\end{align}
for hopping terms that do not cross the boundary. 
For convenience, let
us introduce two sets of natural numbers, $4\mathbb{N}=\{4n \ | n\in\mathbb{N}\}$ and
$4\mathbb{N}+2=\{4n+2 \ | n\in\mathbb{N}\}$, where $\mathbb{N}$ is the set of natural
numbers. 
For the hopping term across the boundary, the string operator in the JWT becomes
trivial when the system size satisfies $L\in 4\mathbb{N}$ for APBCs or $L\in4\mathbb{N}+2$
for PBCs. This corresponds to a closed-shell condition, where a gap always persists at
half filling, except at the critical point $v=w$. Thus, the SSH model can be exactly
mapped onto the following qubit model:
\begin{align}
\hat{\mathcal{H}}_{\text{SSH}} = - \sum_{ i=1 }^{L/2} \left(  v \sigma_{A,i}^+ \sigma_{B,i}^-  + w \sigma_{B,i}^+ \sigma_{A,i+1}^- \right)  + \text{H.c.}
\label{eq:H_SSH_JWT}
\end{align}
The total Hamiltonian can be decomposed into two groups, each containing mutually
commutable local Hamiltonian terms, i.e.,
\begin{align}
\hat{\mathcal{H}}_{\text{SSH}} = \hat{\mathcal{H}}_1 +  \hat{\mathcal{H}}_2,
\end{align}
where
\begin{align}
\hat{\mathcal{H}}_1 = - \sum_{i = 1}^{L/2} v \sigma_{A,i}^+ \sigma_{B,i}^-  + \text{H.c.}
\end{align}
and
\begin{align}
\hat{\mathcal{H}}_2 = - \sum_{i=1}^{L/2}  \gamma w \sigma_{B,i}^+ \sigma_{A,i+1}^-   + \text{H.c.}.
\end{align}
Each term in $\hat{\mathcal{H}}_i~(i=1, 2)$ involves interactions between only two
neighboring qubits. The ground state of an individual term
$-(\sigma_{i}^{+}\sigma_{i+1}^{-} + \text{H.c.})$ is simply given by
$|{\rm t}\rangle = \frac{1}{\sqrt{2}}\left(|01\rangle + |10\rangle\right)$, which can be
easily prepared from the product state $|00\rangle$ using the circuit shown in
Fig.~\ref{fig:fig2}(a). Therefore, the ground state of $\mathcal{H}_i~(i=1, 2)$ can be
prepared by applying $L/2$ identical copies of this set of gates to neighboring qubits.
We choose $\hat{\mathcal{H}}_i~(i=1, 2)$ as our initial Hamiltonian.

Throughout this paper, we examine the unitary evolution from the ground state of
$\hat{\cal {H}}_1$ to that of $\hat{\cal {H}}_{\text{SSH}}$ using the DQAP
ansatz~\cite{PhysRevResearch.3.013004}:
\begin{eqnarray}
|\Psi_M({\bm \theta})\rangle &=& \prod_{m=M}^1
\hat{\mathcal{U}}^{(1)}({\theta_m^{(1)},\theta_m^{(2)}}) |\Psi^{(1)}\rangle
\nonumber \\
&=& \hat{\mathcal{U}}^{(1)}({\theta_M^{(1)},\theta_M^{(2)}})\cdots
\hat{\mathcal{U}}^{(1)}({\theta_2^{(1)},\theta_1^{(2)}})
\hat{\mathcal{U}}^{(1)}({\theta_1^{(1)},\theta_1^{(2)}}) |\Psi^{(1)}\rangle,
\label{eq:DQAP}
\end{eqnarray}
where
\begin{align}
\hat{\mathcal{U}}^{(1)}({\theta_m^{(1)},\theta_m^{(2)}}) =
e^{ - i \theta_{m}^{(1)} \hat{\mathcal{H}}_1 }  e^{ - i \theta_{m}^{(2)} \hat{\mathcal{H}}_2 }
\label{eq:DQAP_U}
\end{align}
and $M$ represents the number of time steps with
$|\Psi^{(1)}\rangle = |{\rm b}\rangle_1 \otimes |{\rm b}\rangle_2 \otimes \cdots \otimes |{\rm b}\rangle_{L/2} $
being the ground state of $\hat{\mathcal H}_1$. The variational parameters
$\{\theta_{m}^{(1)}, \theta_{m}^{(2)}\}_{m=1}^M$ are optimized by minimizing the
expectation value of the energy:
\begin{equation}
E_M({\bm \theta}) = \langle \Psi_{M}({\bm \theta})|\hat{\cal H}_{\text{SSH}}|
\Psi_M({\bm \theta}) \rangle.
\label{eq:energy}
\end{equation}
We employ the natural gradient descent method for optimization, where both the gradient
and the metric tensor are efficiently calculated as described in
Ref.~\cite{PhysRevResearch.3.013004}. Since $\hat{\mathcal{H}}_i~(i=1, 2)$ consists of
mutually commutative terms, the time evolution of $\hat{\mathcal{H}}_i~(i=1, 2)$ can be
efficiently parallelized. The time-evolution of the local term
$\exp[- i \theta (\sigma_i^{+} \sigma_{i+1}^{-} + \text{H.c.})]$ can be exactly
implemented using the circuit shown in Fig.~\ref{fig:fig2}(b).

 Finally, we note that the DQAP ansatz in
Eqs.~(\ref{eq:DQAP}-\ref{eq:energy}) at an intermediate step generally does not represent
the ground state of the instantaneous Hamiltonian~\cite{PhysRevResearch.3.013004}
somewhere on the straight line of a Path connecting the start and end points in
Fig.~\ref{fig:fig1} (b). Therefore, the straight lines connecting the start and end points
in Fig.~\ref{fig:fig1} (b) should be understood as schematics.

\section{Numerical Results} \label{sec:results}

In this section, we present numerical results obtained using classical computers. The
initial state is always set to the ground state of $\hat{\mathcal{H}}_1$ with $v=1$, which
is equivalent to the ground state of $\hat{\mathcal{H}}_{\rm SSH}$ with $(v, w)=(1, 0)$.
We then vary the parameters $(v, w)$ in the final Hamiltonian
$\hat{\mathcal{H}}_{\rm SSH}$, whose ground state is the target final state in the time
evolution [see Fig.~\ref{fig:fig1}(b)]. By appropriately tuning $(v, w)$ in the final
Hamiltonian, we systematically investigate the two distinct cases: one where the time
evolved state remains within the same topologically trivial phase, and another where it
transforms into a topologically nontrivial phase. The main results are summarized in
Table~\ref{tab}.

\subsection{Ground-state energy}

\begin{figure}
\includegraphics[width=0.95\columnwidth]{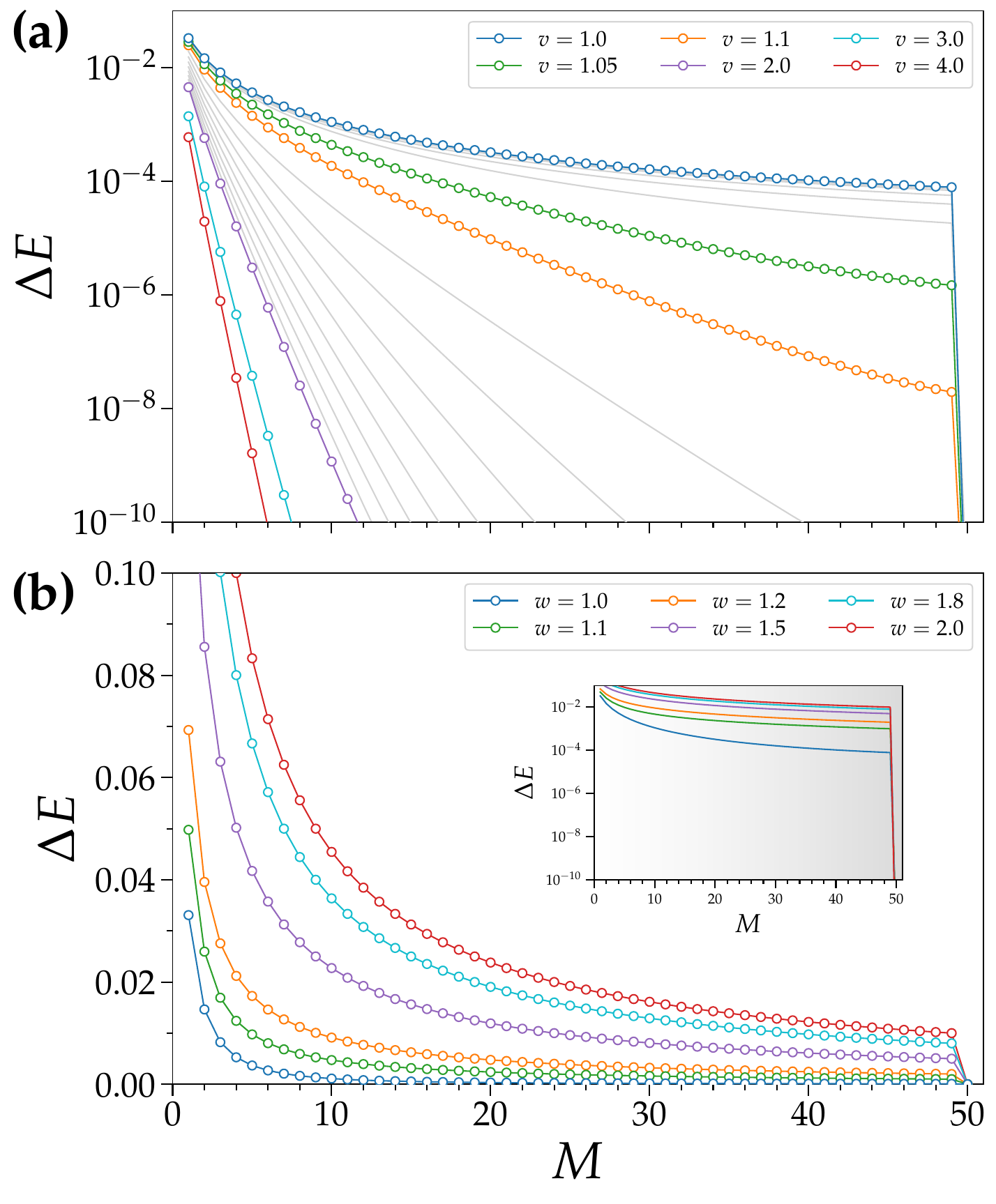}
\caption{The energy difference $\Delta E$ as a function of circuit depth $M$ for various
 parameters $(v, w)$ in the final Hamiltonian. The system size is $L = 200$ under APBCs.
 (a) $w$ is fixed at $1$, while $v$ is varied, representing the case where both the initial
 and final states belong to the trivial phase (case I). (b) $v$ is fixed at $1$, while $w$
 is varied, illustrating the case where the initial and final states belong to different
 topological phases (case II). In (a), thin lines between the results for $v=1.0$ and
 $v=1.05$ indicate intermediate values from $v=1.01$ to $v=1.04$, while thin lines between
 $v=1.1$ and $v=2$ correspond to $v=1.2$ to $v=1.9$. The inset of (b) displays a
 semi-logarithmic plot.
}
\label{fig:fig3}
\end{figure}

Figure~\ref{fig:fig3} shows the energy difference,
$\Delta E = E_L^{M}({\bm \theta}_{\text{opt}}) - E_{\text{exact}}(L)$, between the
optimized variational ground-state energy $E_L^{M}({\bm \theta}_{\text{opt}})$ and the
exact ground-state energy $E_{\text{exact}}(L)$ for a system size of $L=200$. Here,
$E_L^{M}({\bm \theta}_{\text{opt}})$ is computed using the DQAP ansatz
$|\Psi_M({\bm \theta})\rangle $ with the optimized variational parameters
${\bm \theta}=\{\theta_m^{(1)}, \theta_m^{(2)}\}_{m=1}^M$, which defines a quantum circuit
with a depth of $M$. The optimized variational parameters ${\bm \theta}_{\text{opt}}$ for
different values of $M$ are provided in Appendix~\ref{app:opt} (see also
Figs.~\ref{fig:fig14} and~\ref{fig:fig15}). In Fig.~\ref{fig:fig3}(a), $w$ is fixed at
$1$, while $v$ is varied form $1$ to 4 in the final Hamiltonian. In contrast, in
Fig.~\ref{fig:fig3}(b), $v$ is fixed at $1$ and $w$ is varied from $1$ to $2$. These
choices correspond to different time evolution scenarios: Fig.~\ref{fig:fig3}(a)
represents an evolution where the system remains in the trivial phase (Path I), while
Fig.~\ref{fig:fig3}(b) corresponds to a transition from the trivial to the topological
phase (Path II). It is important to note that at the critical point $v=w=1$ in the final
Hamiltonian, the exact ground state can be prepared when $M = L/4$ for APBCs, as shown in
Fig.~\ref{fig:fig3}, or $M=(L-2)/4$ for PBCs~\cite{PhysRevResearch.3.013004}. This result
is consistent with the bound imposed by the information propagation speed in a quantum
circuit composed of local quantum gates~\cite{LiebEH1972}. The system size dependence of
the energy is discussed in Appendix~\ref{app:en}.

In addition, we observe three notable features in Fig.~\ref{fig:fig3}. First, as shown in
Fig~\ref{fig:fig3}(a), within the parameter range $v\in[1.0, 1.1]$ in the final
Hamiltonian, the number of layers required to prepare the ground state with extremely high
accuracy ($< 10^{-8}$) remains a quarter of the system size. This is reasonable since this
parameter range is very close to the critical point. Second, as $v$ increases further,
$\Delta E$ exhibits a clear exponential decay with respect to $M$. This implies that a
shallow circuit, containing fewer quantum gates than in the critical case ($v=1$), is
sufficient to prepare the ground state with high accuracy. This observation holds when the
initial and final states belong to the same phase and when a large spectrum gap persists.
A similar conclusion is expected for cases where both the initial and final states reside
in the non-trivial phase, as these cases can be mapped onto each other by a single-site
translation.
Third, as shown in Fig.~\ref{fig:fig3}(b), when the initial and final states belong to
different topological phases, the minimal number of layers required to exactly prepare the
target ground state is consistently $M = L/4$.

\subsection{Entanglement entropy}

\begin{figure}
\includegraphics[width=0.95\columnwidth]{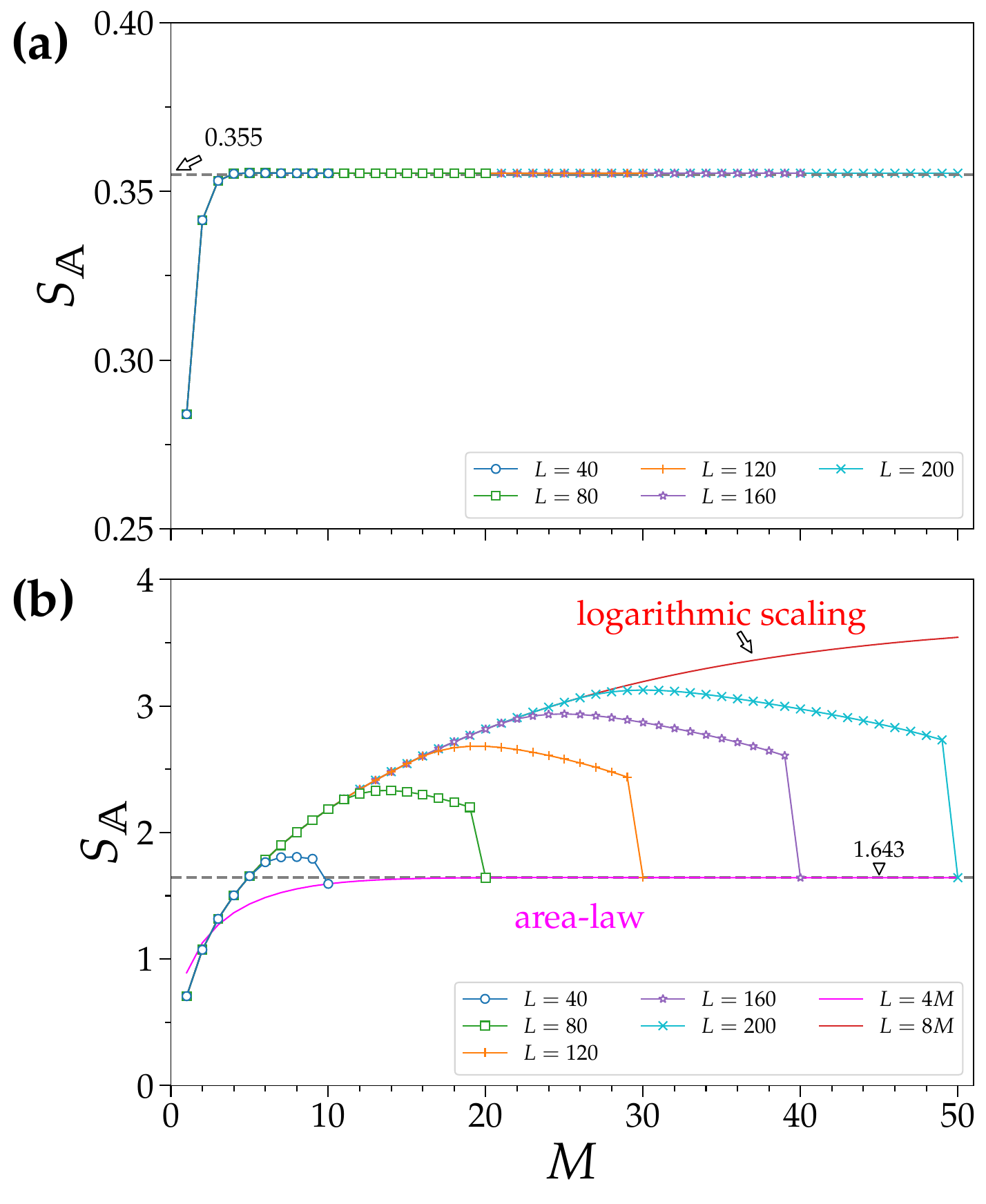}
\caption{Entanglement entropy $S_{\mathbb A}$ as a function of circuit depth $M$ for
various system sizes $L$ under APBCs. (a) and (b) show the results for the parameters
$(v, w)=(2.0, 1.0)$ and $(1.0, 1.1)$ in the final Hamiltonian, respectively,
corresponding to Path I and Path II in Fig.~\ref{fig:fig1}(a). The variational parameters
$\bm\theta$ in the DQAP ansatz $|\Psi_M(\bm\theta)\rangle$ are optimized for each $L$ and
$M$. For comparison, the results for the case where $L$ and $M$ are varied while
maintaining $L=8M$ are also shown by a read line. Additionally, the entanglement entropy
for the exact ground state of the final Hamiltonian with the system size $L$ is plotted at
$L=4M$ using a magenta line.}
\label{fig:fig4}
\end{figure}

In Fig.~\ref{fig:fig4}, we calculate the evolution of the entanglement entropy,
$S_{\mathbb A} = -\text{Tr}\rho_{\mathbb A} \ln \rho_{\mathbb A} $, for two different
cases. Here,
$\rho_{\mathbb A}=\text{Tr}_{\overline{\mathbb A}}|\Psi_M({\bm \theta}_{\rm opt})\rangle \langle \Psi_M({\bm \theta}_{\rm opt})|$
is the reduced density matrix of the optimized state
$|\Psi_M({\bm \theta}_{\rm opt})\rangle$ for subsystem $\mathbb A$. We consider a
half-system bipartition into contiguous subsystems $\mathbb A$ and its complement
$\overline{\mathbb A}$, placing the entanglement cuts on $w$-bonds (i.e., bonds between
unit cells). Details of the calculation can be found in
Ref.~\cite{PhysRevResearch.3.013004}.
In Fig.~\ref{fig:fig4}(a), where both the initial and final states belong to the trivial
phase (Path I), we observe that $S_{\mathbb A}$ saturates at approximately $0.355$ after only four layers. This indicates that in the gapped case, the two halves of
the system are weakly entangled, and a shallow circuit of depth $M=4$ is sufficient to
generate the necessary quantum entanglement between subsystems $\mathbb A$ and
$\overline{\mathbb A}$ in the target state.

\begin{figure}
\includegraphics[width=0.95\columnwidth]{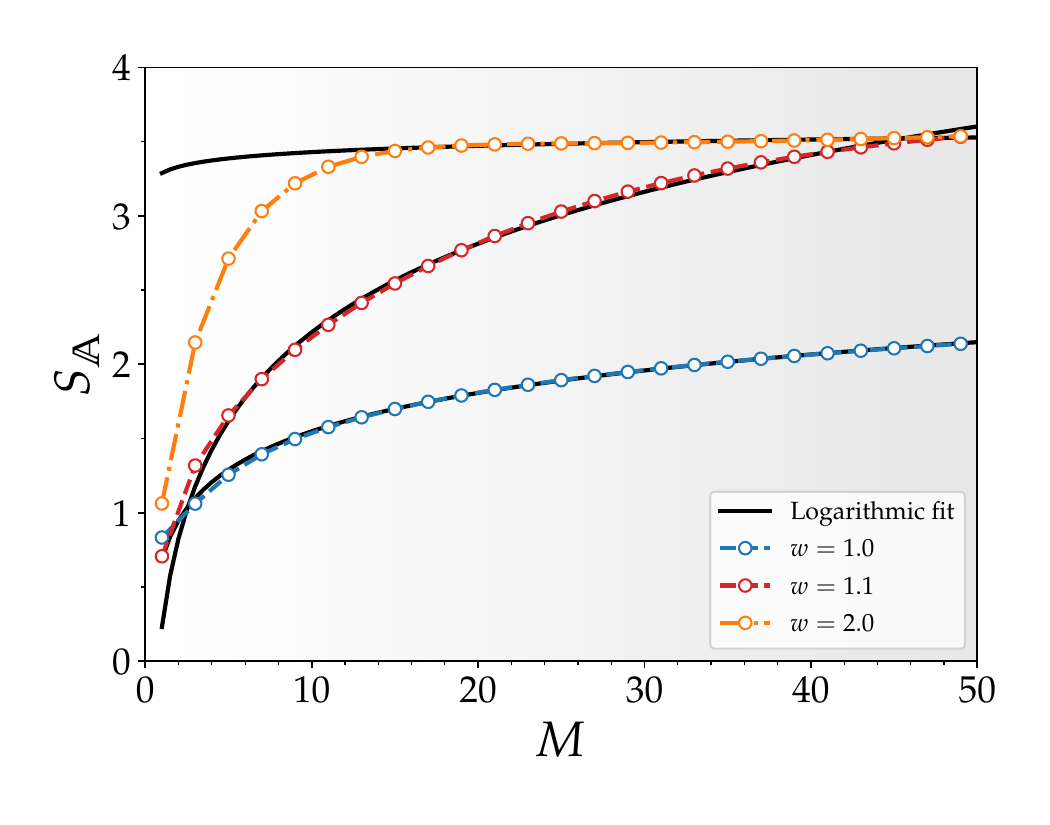}
\caption{Logarithmic fitting (solid lines) of the entanglement 
entropy 
$S_{\mathbb A}$ for $M=L/8$ (circles), obtained using the 
optimzed DQAP ansatze $|\Psi_M(\bm\theta)\rangle$ for various system 
sizes $L$.
The fitting parameters $(a,b)$ for the form 
$\tilde{S}_{\mathbb A} = a \ln{M} + b$ are 
$(0.3709, 0.6942)$, 
$(0.8619, 0.2300)$, and
$(0.0612, 3.2880)$ 
for the parameter sets $(v,w) = (1.0, 1.0)$, $(1.0, 1.1)$ and 
$(1.0, 2.0)$, respectively.}
\label{fig:fig5}
\end{figure}

In Fig.~\ref{fig:fig4}(b), where the initial and final states belong to different
topological phases (Path II), the entanglement entropy $S_{\mathbb A}$ exhibits a clearly
nonmonotonic dependence on the number $M$ of circuit layers. We note that such
nonmonotonic behavior was not observed clearly in the previous study at the critical point
$v=w=1$~\cite{PhysRevResearch.3.013004}. Upon closer examination of its dependence on $M$,
we find that $S_{\mathbb A}$ initially increases, following a universal, system-size
independent curve for $M < L/8$. This behavior can be understood in terms of operator
spreading: two-qubit gates acting on the two boundaries between $\mathbb A$ and
$\overline{\mathbb A}$ do not overlap in the DQAP ansatz as long as
$M < L/8$~\cite{PhysRevResearch.3.013004}. Specifically, the entanglement between
$\mathbb A$ and $\overline{\mathbb A}$ is first generated by two-qubit gates on the
boundaries at $M=1$, and for $M < L/8$, the supports of these operators remain
non-overlapping, leading to a system-size independent $S_{\mathbb A}$.

To gain insights into the universal behavior of the entanglement 
entropy $S_{\mathbb A}$ for $M=L/8$, we fit the data obtained 
for various system sizes $L$ using the following 
form~\cite{Jobst2022}:  
\begin{equation}
\tilde{S}_{\mathbb A} = a \ln{M} + b,  
\end{equation}
where $a$ and $b$ are fitting parameters. 
Figure~\ref{fig:fig5} shows the fitting results for the 
parameter sets $(v,w)=(1.0,1.0), (1.0, 1.1)$, and $(1.0,2.0)$.  
We find that the entanglement entropy $S_{\mathbb A}$ fits well 
to a logarithm form for larger $M$, even when the target ground 
state is not at criticality.

At $M=L/8$, $S_{\mathbb A}$ begins to deviate from the universal curve and, almost
simultaneously, starts decreasing as $M$ increases.
$S_{\mathbb A}$ finally reaches the exact value of the final Hamiltonian
$\hat{\cal H}_{\rm SSH}$ at $M=L/4$, exhibiting a discontinuity.
Notably, the system-size dependence of the exact values is insignificant, as indicated by
the magenta line in Fig.~\ref{fig:fig4}(b).
This implies that the entanglement entropy $S_{\mathbb A}$ of the 
ground state at $(v,w)=(1.0,1.1)$ follows the area-law scaling, 
in contrast to the critical case, where the
entanglement entropy increases logarithmically with the subsystem
size~\cite{Calabrese2004}.
While the convergence of $S_{\mathbb A}$ to the exact value at $M=L/4$ is expected, the
observed clear discontinuity as a function of circuit depth $M$ was not found in the
previous study for the critical case~\cite{PhysRevResearch.3.013004}. These finding
suggests a nontrivial information spreading process during the unitary evolution from the
trivial to the topological phase under the DQAP ansatz.

\subsection{Mutual information} \label{sec:mi}

\begin{figure}
\includegraphics[width=0.95\columnwidth]{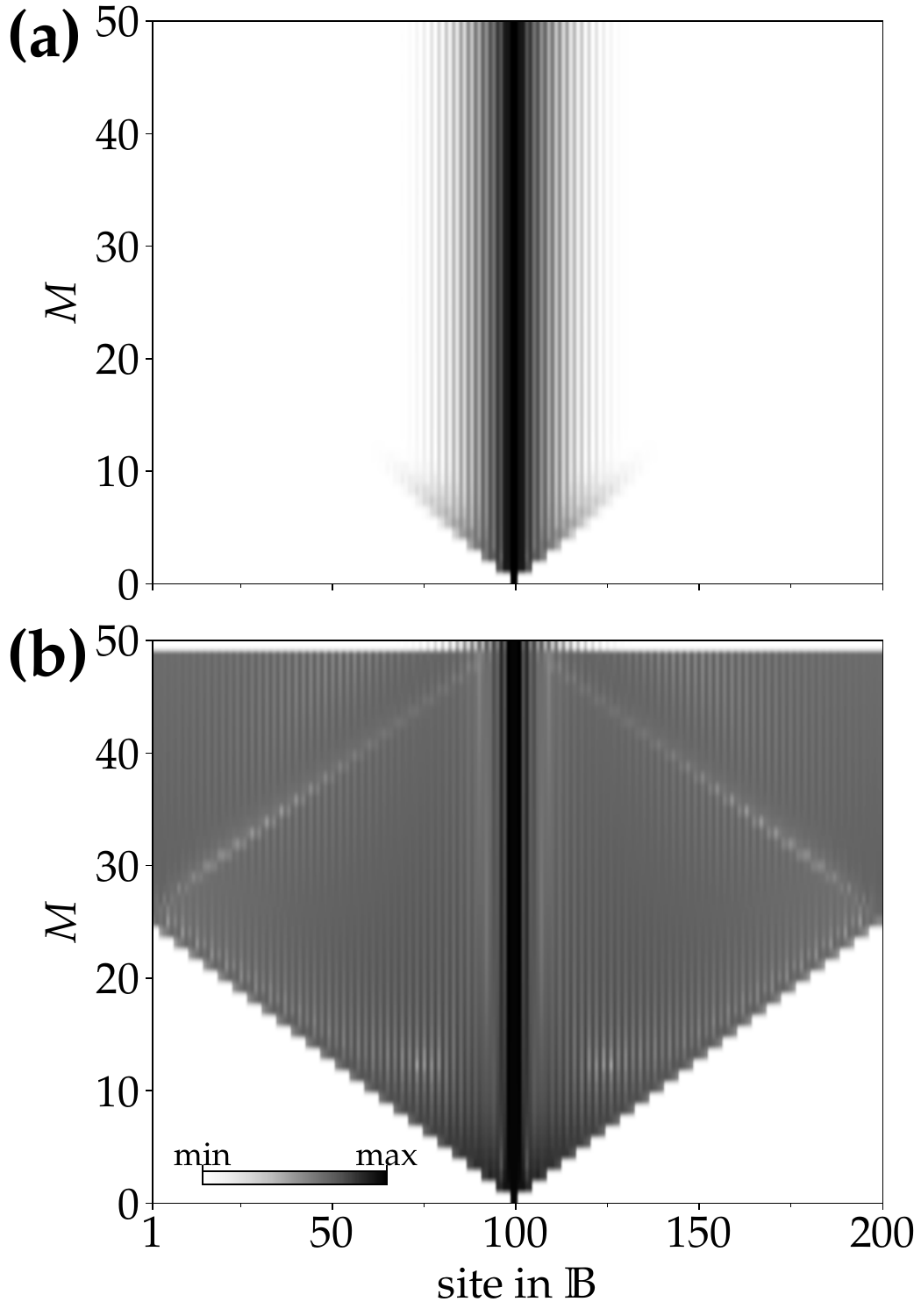}
\caption{Intensity plot of the mutual information $I_{\mathbb{A}, \mathbb{B}}$ for the
optimized DQAP ansatz $|\Psi_M(\bm\theta)\rangle$, shown as a function of the circuit
depth $M$ and the location of the single site composing subsystem $\mathbb B$. (a) and (b)
correspond to the cases with the parameters $(v, w)=(2, 1)$ and $(1, 2)$ in the final
Hamiltonian, respectively, representing Path I and Path II in Fig.~\ref{fig:fig1}(a). The
system size is set to $L=200$ under APBCs and the DQAP ansatz $|\Psi_M(\bm\theta)\rangle$
is optimized for each $M$. Subsystem $\mathbb{A}$ consists of two sites locating at the
center of the system. White color in the intensity plots represent zero. 
}
\label{fig:fig6}
\end{figure}

To further examine the information spreading process under the DQAP ansatz, we compute
the mutual information.
The mutual information between two parts of the system is defined as
$I_{\mathbb{A}, \mathbb{B}} = S_{\mathbb{A}} + S_{\mathbb{B}} - S_{\mathbb{A}\cup \mathbb{B}}$,
which quantities the entanglement between the two parts $\mathbb{A}$ and $\mathbb{B}$ of
the system.
Fixing the system size at $L=200$, we define subsystem $\mathbb A$ as two sites at the
99th and 100th sites (i.e., the 50th unit cell, corresponding to a $v$-bond), while
subsystem $\mathbb{B}$ consists of a single site, whose location is varied.
Figure~\ref{fig:fig6} shows the mutual information for Path I and Path II.
For Path I, the mutual information remains spatially localized around the reference
subsystem $\mathbb{A}$ throughout the entire evolution process.
In contrast, for Path II, the mutual information gradually spreads across the system and
extends throughout the whole system by the time the circuit depth reaches $M=L/8$. It then
remains spread across the entire system until $M=L/4-1$, just one layer before reaching
the exact ground state of the target Hamiltonian $\hat{\cal H}_{\rm SSH}$ with
$(v, w)=(1, 2)$. Finally, at $M=L/4$, the mutual information abruptly localized around the
reference subsystem $\mathbb{A}$. This sudden change of mutual information at the final
layer is compatible with the discontinuous behavior of the entanglement entropy observed
in Fig.~\ref{fig:fig4}(b). A further analysis of the mutual information can be found in
Appendix~\ref{app:mi2}.

\subsection{Polarization}

Having observed the nontrivial information-spreading process along Path II under the DQAP
ansatz, a natural question arises: At what circuit depth $M$ does the ansatz state
$|\Psi_M({\bm \theta})\rangle$ undergo a topological transition? To address this, we
compute the polarization defined as~\cite{PhysRevLett.80.1800, PhysRevX.8.021065}
\begin{equation}
{\cal P}_{\rm R}(M) =  {\rm Im} \ln \langle \Psi_M({\bm \theta}_{\rm opt})|\hat{\cal U}_{\rm R} |\Psi_M({\bm \theta}_{\rm opt}) \rangle,
\label{eq:pol_orig}
\end{equation}
where the unitary operator $\hat{\cal U}_{\rm R}$ is defined as
\begin{align}
\hat{\cal U}_{\rm R}
&={\exp}\left[\frac{2\pi i}{(L/2)} \sum_{j=1}^{L/2}
\left(
\left(j-\frac{1}{2}\right)  \hat{c}_{A,j}^\dag \hat{c}_{A,j}
+ j \hat{c}_{B,j}^\dag \hat{c}_{B,j}\right)
\right] \notag \\
&={\exp}\left[\frac{2\pi i}{L} \sum_{l=1}^{L}
l \hat{c}_{l}^\dag \hat{c}_{l}
\right]
\notag \\
&=\prod_{l=1}^{L}{\exp}\left[\frac{2\pi i}{L}
l \hat{c}_{l}^\dag \hat{c}_{l}
\right].
\label{eq:UR}
\end{align}
Here, $l$ represents the site index, serving as a one-dimensional label that runs over
both unit-cell and sublattice indexes $j$, $A$, and $B$. Specifically, we define
$l = 2j-1$ ($l=2j$) for sites belonging to sublattice $A$ ($B$) of the $j$th unit cell.
The last equality in Eq.~(\ref{eq:UR}) holds because fermion density operators commute
with each other.

\begin{figure}
\includegraphics[width=0.95\columnwidth]{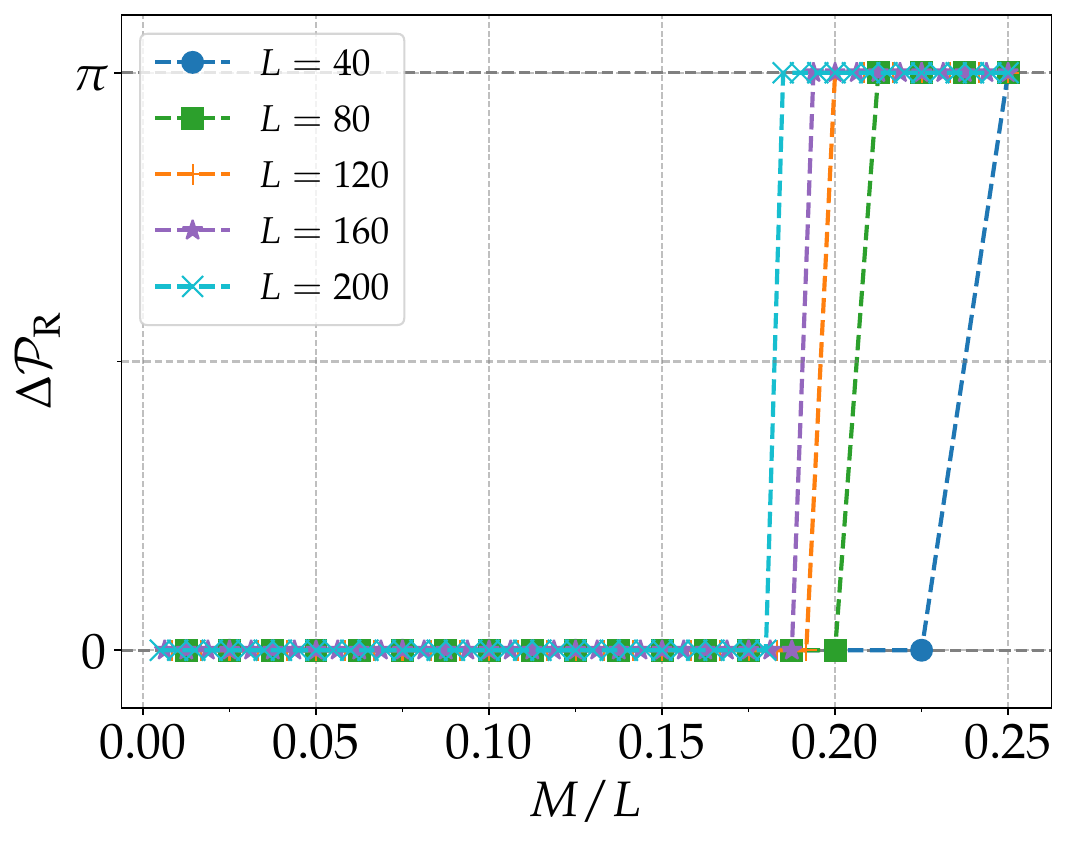}
\caption{The polarization difference $\Delta{\cal P}_{\rm R}(M)$ as a function of the
circuit depth $M$ for various system sizes $L$. The variational parameters $\bm\theta$ in
the DQAP ansatz $|\Psi_M(\bm\theta)\rangle$ are optimized for each $M$ and $L$. The
parameters in the final Hamiltonian are set to $(v, w)=(1.0, 1.1)$,
corresponding to Path II in Fig.~\ref{fig:fig1}(b). 
}
\label{fig:fig7}
\end{figure}

The polarization ${\cal P}_{\rm R}(M)$ depends on the choice of the origin for the
``position" $l$ in front of $\hat{c}_l^\dag \hat{c}_l$ in Eq.~(\ref{eq:UR}). However, the
difference
\begin{align}
\Delta {\cal P}_{\rm R}(M) = {\cal P}_{\rm R}(M) -  {\cal P}_{\rm R}(0)
\end{align}
allows for an unambiguous detection of the topological phase transition during the DQAP
evolution. Figure~\ref{fig:fig7} shows the calculated polarization as a function of the
circuit depth $M$ for various system sizes $L\in 4\mathbb{N}$ along Path II. As expected,
the polarization takes different values in the initial $(M=0)$ and final $(M=L/4)$ states,
indicating the occurrence of a topological phase transition during the DQAP evolution.

\begin{figure}
\includegraphics[width=0.95\columnwidth]{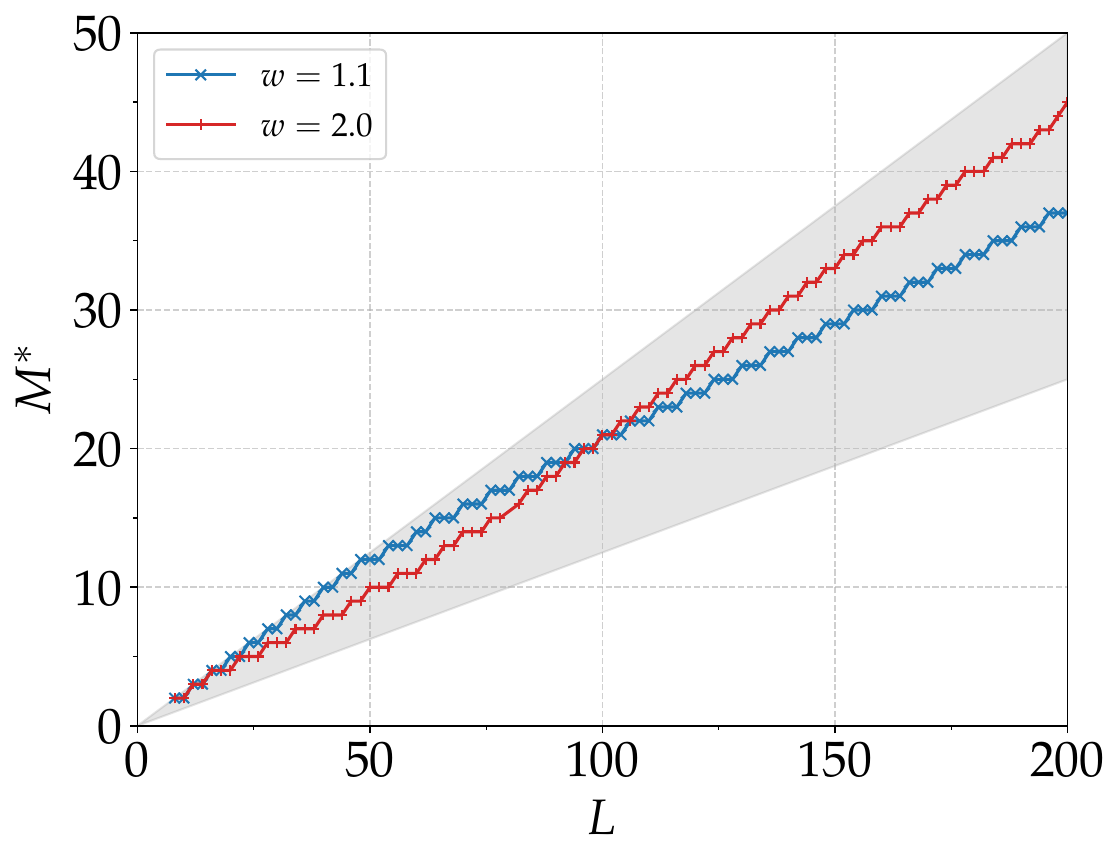}
\caption{The critical circuit depth $M^*$ as a function of the system size $L$. The
parameters of the final Hamiltonian are set to $(v, w)=(1.0, 1.1)$ and $(1.0, 2.0)$. APBCs are
imposed for $L\in4\mathbb{N}$, while PBCs are used for $L\in4\mathbb{N}+2$.
The shaded region indicates $L/8\leqslant M^* \leqslant L/4$.
}
\label{fig:fig8}
\end{figure}

Interestingly, the critical circuit depth $M^*$ at which ${\cal P}_R(M)$ changes,
\begin{equation}
{\cal P}_{\rm R}(M^*) -  {\cal P}_{\rm R}(M^*-1)\ne 0,
\label{eq:mstar}
\end{equation}
does not generally coincide with the final step. This suggests that the topological phase
transition occurs during the DQAP evolution, before reaching the exact ground state of the
final Hamiltonian. Figure~\ref{fig:fig8} shows $M^*$ as a function of the the system size
$L$ for two different parameter sets, $(v, w)=(1, 1.1)$ and $(1, 2)$, in the final
Hamiltonian. We observe that $M^*$ exhibits a staircase-like increase with $L$. Although
it is not conclusive, our numerical results suggest that the critical $M^*$ satisfies
$L/8 \leqslant M^* \leqslant L/4$ for $L \in 4\mathbb{N}$ and
$(L-2)/8 \leqslant M^* \leqslant (L-2)/4$ for $L \in 4\mathbb{N}+2$, where the lower bound
corresponds to the circuit depth at which the causal cone spans the entire system [see
Fig.~\ref{fig:fig6}(b) and Ref.~\cite{PhysRevResearch.3.013004}]. Remarkably, if the
topological phase transition occurs before the final step, no discontinuities in
ground-state energy, entanglement entropy, or mutual information are observed at $M=M^*$.

\section{Experiments on quantum hardware} \label{sec:qc}

Exploring topological phase transitions poses a significant challenge for near-term noisy
quantum devices, as topological order parameters are inherently nonlocal, as seen in
Eq.~(\ref{eq:UR}). Nevertheless, a seminal study~\cite{Smith2022} successfully
demonstrated a topological phase transition in a spin chain by evaluating topological
order parameters using IBM's superconducting quantum computers. Here, in this work, we
employ Quantinuum's trapped-ion quantum computer to evaluate the polarization
${\cal P}_{\rm R}$ as a topological order parameter. We further detail how the abrupt
change in polarization can be detected with a real quantum device.

The experiments were conducted in January 2024 using the H1-1 system by
Quantinuum~\cite{H1datasheet}. At the time of the experiments, the H1-1 system consisted
of $20$ qubits and natively supported single-qubit rotation gates and two-qubit ZZ phase
gates defined as
${\rm ZZPhase}(\alpha) = e^{-\frac{1}{2} i \pi \alpha \hat{Z}_i \hat{Z}_j}$, parametrized
by a real angle~$\alpha$. These native two-qubit native gates could be applied between an
arbitrary pair of qubits. The average infidelity of single-qubit and two-qubit gates was
about $0.004\%$ and $0.2\%$, respectively, while state preparation and measurement errors
averaged $0.3\%$. Further details on the hardware specifications can be found in
Ref.~\cite{H1datasheet}. All quantum circuits used in the experiments were compiled using
TKET~\cite{Sivarajah_2020}.

\begin{table}
\caption{The optimized variational parameters
$\bm\theta_{\rm opt} =\{\theta_m^{(1)}, \theta_m^{(2)} \}_{m=1}^M$
obtained from classical simulations for $L=18$ under PBCs,
which are subsequently used for the experiments.
\label{tab:opt}
}
\begin{tabular}{c|cccc}
\hline \hline
& $M=1$ & $M=2$ & $M=3$ & $M=4$ \\
\hline
$\theta_1^{(1)}$ & 0.7853981636  & 1.2494387001 & 1.3583873392 & 1.4379338692 \\
$\theta_1^{(2)}$ & 0.2767871793  & 0.2688075377 & 0.2534789267 & 0.5229341500 \\
$\theta_2^{(1)}$ & -- & 0.6392420907 & 1.1586546608 & 1.4498686393 \\
$\theta_2^{(2)}$ & -- & 0.4831535535 & 0.5146221144 & 0.7033476173 \\
$\theta_3^{(1)}$ & -- & -- & 0.5714210664 & 1.4215754149 \\
$\theta_3^{(2)}$ & -- & -- & 0.5376954104 & 0.7169480772 \\
$\theta_4^{(1)}$ & -- & -- & -- & 1.0837464017 \\
$\theta_4^{(2)}$ & -- & -- & -- & 0.6676928002 \\
\hline \hline
\end{tabular}
\end{table}

We consider the 1D SSH model of $L=18$ sites under PBCs, setting the parameters
$(v, w)=(1, 2)$ in the final target Hamiltonian. Every single site is mapped to a single
qubit via the JWT, as described in Sec.~\ref{sec:model}. Additionally, we introduce an
ancillary qubit for the Hadamard test, bringing the total number of qubits used to
$L+1=19$. The number of circuit layers is varied within the range
$0 \leqslant M \leqslant 4=(L-2)/4$. To optimize quantum resource utilization, the optimal
variational parameters $\boldsymbol{\theta}_{\rm opt}$ are pre-determined through
classical simulations (see Table~\ref{tab:opt}), allowing us to bypass the iterative
parameter optimization process between quantum and classical computers.

The polarization given in Eq.~(\ref{eq:pol_orig}) can be reformulated for quantum
computation as
\begin{equation}
{\cal P}_{\rm R}(M) = \arctan{\left(\frac{y}{x}\right)}
\label{eq:pol2}
\end{equation}
with
\begin{align}
x&={\rm Re}\langle \Psi_M(\bm \theta_{\rm opt}) |\hat{\cal U}_{\rm R}|\Psi_M(\bm \theta_{\rm opt})\rangle,
\label{eq:x} \\
y&={\rm Im}\langle \Psi_M(\bm \theta_{\rm opt}) |\hat{\cal U}_{\rm R}|\Psi_M(\bm \theta_{\rm opt})\rangle.
\label{eq:y}
\end{align}
We evaluate $x$ and $y$ separately on a quantum computer using the Hadamard test.
The uncertainty in the polarization, arising from the measurement of these
quantities, is estimated using the error propagation formula:
$ \delta {\cal P}_{\rm R} = \sqrt{\left(\frac{\partial {\cal P}_{\rm R}}{\partial x} \delta x\right)^2 + \left(\frac{\partial {\cal P}_{\rm R}}{\partial y} \delta y\right)^2}= \sqrt{\left(\frac{xy}{x^2+y^2}\right)^2 \left(\frac{\delta x^2}{x^2}+ \frac{\delta y^2}{y^2} \right)}, $
where $\delta x$ and $\delta y$ are the standard errors in the measurement of $x$ and
$y$, respectively.

\begin{figure}
\includegraphics[width=1.0\columnwidth]{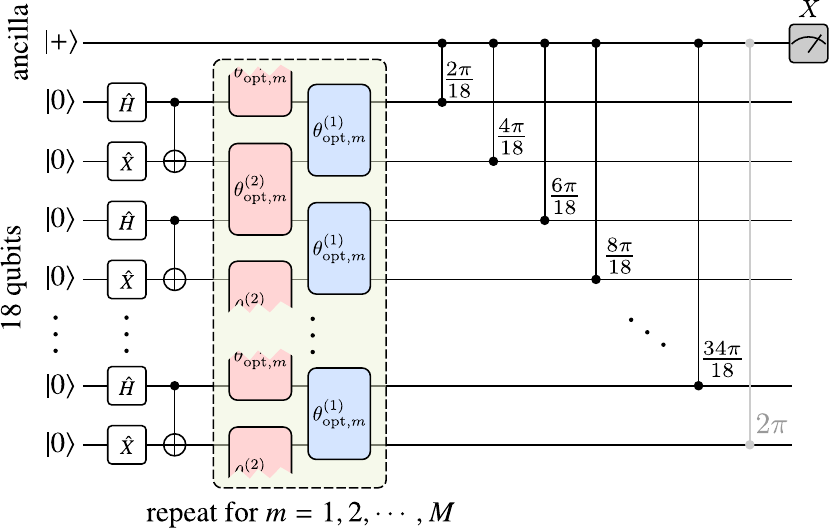}
\caption{The quantum circuit to evaluate the real part
$x={\rm Re}\langle \Psi_M(\bm \theta_{\rm opt}) |\hat{\cal U}_{\rm R}|\Psi_M(\bm \theta_{\rm opt})\rangle$
for $L=18$ under PBCs. The topmost qubit represents the ancillary qubit used for the
Hadamard test, while the remaining 18 qubits correspond to the system qubits, on which the
DQAP state $|\Psi_M({\bm \theta}_{\rm opt})\rangle$ is prepared. The leftmost part of the
circuit, consisting of qubit initialization, Hadamard gates, Pauli $X$ gates, and CNOT
gates, construct the initial state $|\Psi^{(1)}\rangle$, which represents the ground state
of $\hat{\cal H}_1$. The section inside the dashed box represents the DQAP ansatz unitary,
$\prod_{m=M}^1 \hat{\cal U}^{(1)}(\theta_{m}^{(1)}, \theta_{m}^{(2)})$, with the optimized
variational parameters determined from classical simulations (see Table.~\ref{tab:opt}).
Blue and red rounded rectangles denote the two-qubit gates 
$\exp[-i \frac{\theta_{m}^{(1)}v}{2} (\hat{X}_i \hat{X}_j + \hat{Y}_i \hat{Y}_j)]$ and
$\exp[-i \frac{\theta_{m}^{(2)}w}{2} (\hat{X}_i \hat{X}_j + \hat{Y}_i \hat{Y}_j)]$, respectively. 
The sequence of controlled-phase gates corresponds to the ${\rm C}_0{\text-}\hat{\cal U}_{\rm R}$
operation in Eq.~(\ref{eq:C-U}). Notice that the controlled-phase gate
${\rm C}_0{\text -}{\rm Phase}_l(2\pi l/L)$ for $l=L$ (light gray) is the identity operation and is
therefore omitted from the circuit. }
\label{fig:fig9}
\end{figure}

Figure~\ref{fig:fig9} illustrates the quantum circuit to evaluate the real part $x$. First, the state
$|\Psi_M(\bm \theta_{\rm opt})\rangle$ is prepared by applying $M$ layers of the DQAP
ansatz unitary $\prod_{m=M}^1 \hat{\cal U}^{(1)}(\theta_{m}^{(1)}, \theta_{m}^{(2)})$ to
the initial state $|\Psi^{(1)}\rangle$, which is a product of $|{\rm t}\rangle$ and
corresponds to the ground state of $\hat{\cal H}_1$. Second, the
controlled-$\hat{\cal U}_{\rm R}$ operation is implemented as follows:
\begin{equation}
{\rm C}_0{\text -}\hat{\cal U}_{\rm R}
= \prod_{l=1}^{L} {\rm C}_{0}{\text-}{\rm Phase}_{l} \left(\frac{2\pi l}{L} \right),
\label{eq:C-U}
\end{equation}
where ${\rm C}_0{\text-}\hat{\cal U}_{\rm R}$ represents the unitary operation of
$\hat{\cal U}_{\rm R}$ on the system qubits ($1$st to $L$th registers) controlled by an
ancillary qubit ($0$th register).
The controlled-phase gate
${\rm C}_0{\text-}{\rm Phase}_{l}(\theta)={\rm diag}(1, 1, 1, {\rm e}^{i\theta})$ acts
with a control on the $0$th qubit and a target on the $l$th qubit. Since
${\rm C}_{0}{\text-}{\rm Phase}_{l} \left(\frac{2\pi l}{L}\right)$ for $l=L$ is the
identity operation, the implementation of ${\rm C}_0{\text -}\hat{\cal U}_{\rm R}$
requires $L-1$ controlled-phase gates. Finally, the real part $x$ is obtained by measuring
the expectation value of the Pauli $X$ operator on the ancillary qubit. Similarly, the
imaginary part $y$ is determined using the same quantum circuit, except that the ancillary
qubit is measured in the Pauli $Y$ basis instead of $X$. 
Notice that
these quantum circuits are further compiled for the H1-1 system for execution.
The number of native two-qubit gates required on the H1-1 system for various circuit depth $M$ are 26, 62, 98, 134, and 170 for $M=$ 0, 1, 2, 3, and 4, respectively (see
Appendix~\ref{app:native2Qgates}).

Figure~\ref{fig:figReIm}(a) shows the experimentally evaluated results for $x$ and $y$,
as defined in Eqs.~(\ref{eq:x}) and (\ref{eq:y}), for $M=0, 1, 2, 3, $ and $4$ using the
H1-1 system. Each value of $x$ and $y$ was estimated with 500 measurements for
$M=0, 1, 2, 4$ and 2000 measurements for $M=3$ to ensure a sufficiently small
uncertainty $\delta {\cal P}_{\rm R}$. The larger number of
measurements for $M=3$ was necessary because both $x$ and $y$ are closer to zero, leading
to a larger uncertainty in the polarization, which is nothing but the argument of the
complex number $x+iy$. If the number of measurements were set to a similar value, this
would results in an increased statistical uncertainty [see also the expression for
$\delta {\cal P}_{\rm R}$ immediately after Eq.~(\ref{eq:y})]. In general, a greater
number of measurements is required in the vicinity of the topological phase transition to
maintain a given level of precision. Conversely, if the number of measurements remains
fixed, an increase in the uncertainty $\delta {\cal P}_{\rm R}(M)$ can serve as an
indicator of the topological phase transition, i.e., at $M\sim M^*$.
For comparison, we also include noiseless simulation results obtained from the H1-1E
emulator provided by Quantinuum, using the same number of measurements. The experimental
results show qualitative agreement with these noiseless simulations.
For $M\leqslant 3$, the results lie along the negative side of the $y$-axis, implying
that ${\cal P}_{\rm R}(M) = \arctan\left(\frac{y}{x}\right) \approx -\frac{\pi}{2}$. For
$M = 4$, the results shift to the positive side of the $y$-axis, indicating that
${\cal P}_{\rm R}(M)=\arctan\left(\frac{y}{x}\right) \approx \frac{\pi}{2}$.

\begin{figure}
\includegraphics[width=0.95\columnwidth]{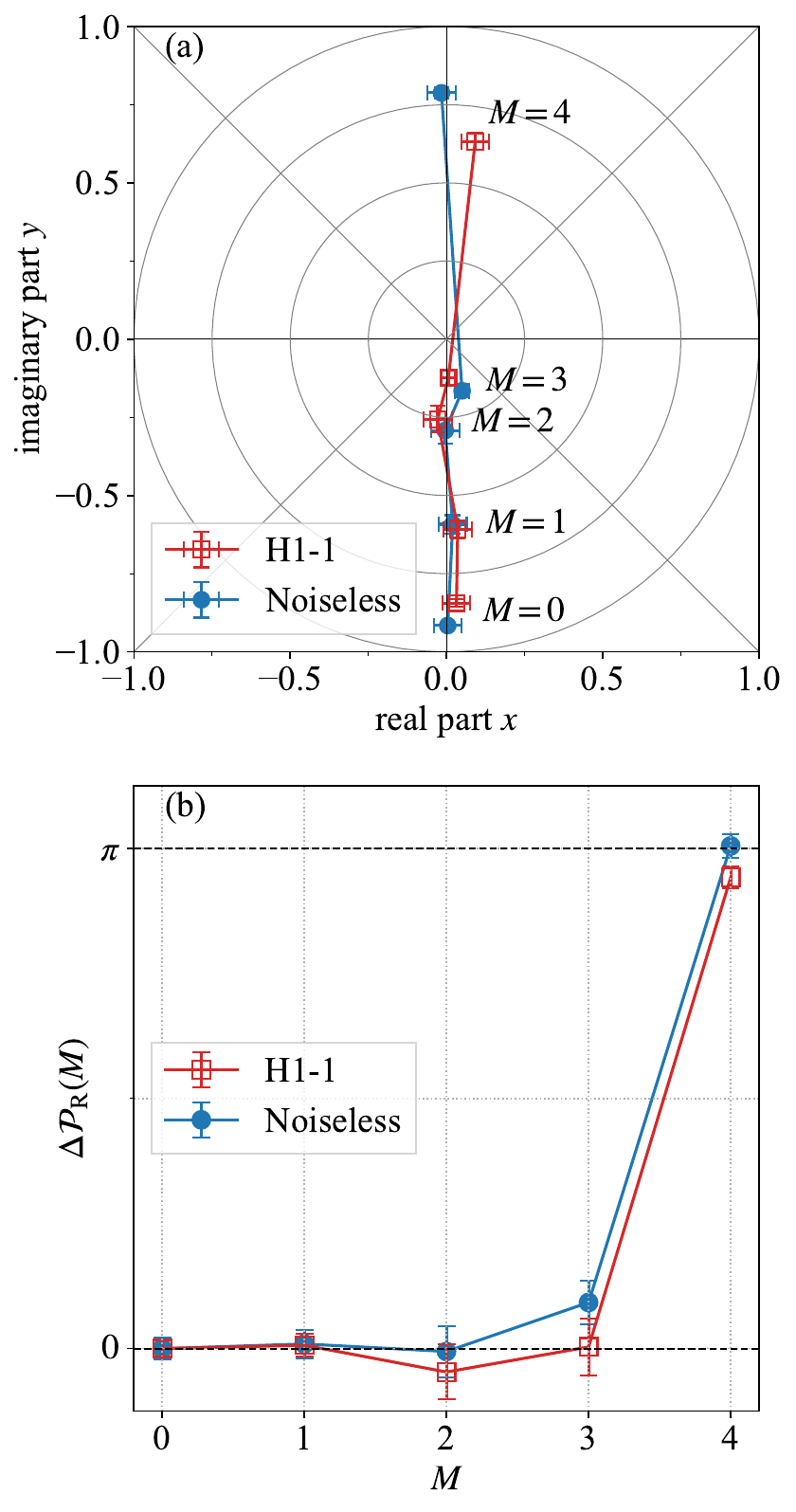}
\caption{
(a)
Real and imaginary parts of $\langle \Psi_M(\bm \theta_{\rm opt}) |\hat{\cal U}_{\rm R}|\Psi_M(\bm \theta_{\rm opt})\rangle$ for $M=0$ to 4 (from bottom to top).
Circular and straight lines are guide to the eyes.
(b)
Polarization difference $\Delta {\cal P}_{\rm R}(M)$ 
as a function of circuit depth $M$. 
Red squares represent the experimental results obtained using 
the H1-1 system, while blue circles denote the noiseless 
simulation results. 
Dashed horizontal lines are guide to the eyes. 
The parameters of the final target Hamiltonian are set to 
$(v,w)=(1,2)$, corresponding to Path II in Fig.~\ref{fig:fig1}(b).
}
\label{fig:figReIm}
\end{figure}

Figure~\ref{fig:figReIm}(b) shows the polarization difference
$\Delta {\cal P}_{\rm R}(M)$, evaluated using the values of $x$ and $y$ shown in
Fig.~\ref{fig:figReIm}. Despite being obtained without any error mitigation, the
experimental results show good quantitative agreement with the noiseless simulations.
Specifically, the polarization at $M=4$ is clearly distinct from those at $M\leqslant 3$.
The robustness of these results against noise can be attributed to the behavior observed
in Fig.~\ref{fig:figReIm}: a small perturbation in $x$ and $y$ does not affect the sign of
$\arctan\left(\frac{y}{x}\right)$. In particular, the substantial change in the
polarization ${\cal P}_{\rm R}(M)$ primarily arises the sign change in $y$ 
(see also additional experimental results in Appendix~\ref{app:experiments}). 
These findings
confirm that our optimized variational state $|\Psi_M({\bm \theta}_{\rm opt})\rangle$
successfully captures the transition between topologically distinct ground states of the
SSH model as the circuit depth $M$ increases, even in the presence of noise in a real
quantum device.

\section{Conclusions} \label{sec:conclusion}

We have applied the DQAP ansatz to obtain the ground state of the 1D SSH model of
spinless fermions at half filling, considering various topological phases to which the
initial and final states belong.
We have found that, irrespective of the topological nature of the initial and final
states, the circuit depth $M$ required to exactly prepare the target final ground state is
$L/4$ for $L\in 4\mathbb{N}$ under APBCs and $(L-2)/4$ for $L \in 4\mathbb{N}+2$ under
PBCs. This is the same circuit depth necessary to obtain the exact ground state at the
critical point when starting from a topologically trivial
phase~\cite{PhysRevResearch.3.013004}.
On the other hand, we have also identified qualitatively distinct behaviors in the
ground-state energy, entanglement entropy, and mutual information during the DQAP
evolution, depending on the topological nature of the initial and final states, as
summarized in Table~\ref{tab}.
One important finding is that as long as the initial and final states belong to the same
phase, only a few layers are sufficient to obtain the target ground state with high
accuracy.
Additionally, we have numerically computed the polarization as an indicator to
distinguish topologically different phases during the DQAP evolution.

We have also demonstrated that the topological phase transition during the DQAP evolution
for the 18-site system can be detected by using a trapped-ion quantum computer. The
all-to-all connectivity of the trapped-ion quantum computer provided by Quantinuum enables
direct evaluation of the polarization, which is derived from the expectation value of a
global unitary operator $\hat{{\cal U}}_{\rm R}$, without introducing additional SWAP
operations. This capability allows us to effectively characterize the phases
experimentally to which the DQAP state belongs.
The present results lay the foundation for the next crucial step--exploring topological
phases in interacting systems~\cite{Hohenadler2013, Rachel2018} using quantum
computers--which we leave for future work.

\section{Achnowledgement}

A part of this work is based on results obtained from project JPNP20017, subsidized by
the New Energy and Industrial Technology Development Organization (NEDO). This study is
also supported by JSPS KAKENHI Grants No. JP21H04446, No. JP22K03479, and No. JP22K03520.
We further acknowledge funding from JST COI-NEXT (Grant No. JPMJPF2221) and the Program
for Promoting Research of the Supercomputer Fugaku (Grant No. MXP1020230411) from MEXT,
Japan. Additionally, we appreciate the support provided by the UTokyo Quantum Initiative,
the RIKEN TRIP initiative (RIKEN Quantum), and the COE research grant in computational
science from Hyogo Prefecture and Kobe City through the Foundation for Computational
Science. The numerical simulations have been performed using the HOKUSAI BigWaterfall
system at RIKEN (Project IDs: Q23604).

\appendix

\section{Additional numerical results} \label{app:results}

In this Appendix, we present additional numerical results for the DQAP ansatz applied to
the 1D SSH model.

\subsection{Energy per site} \label{app:en}

\begin{figure}
\includegraphics[width=0.95\columnwidth]{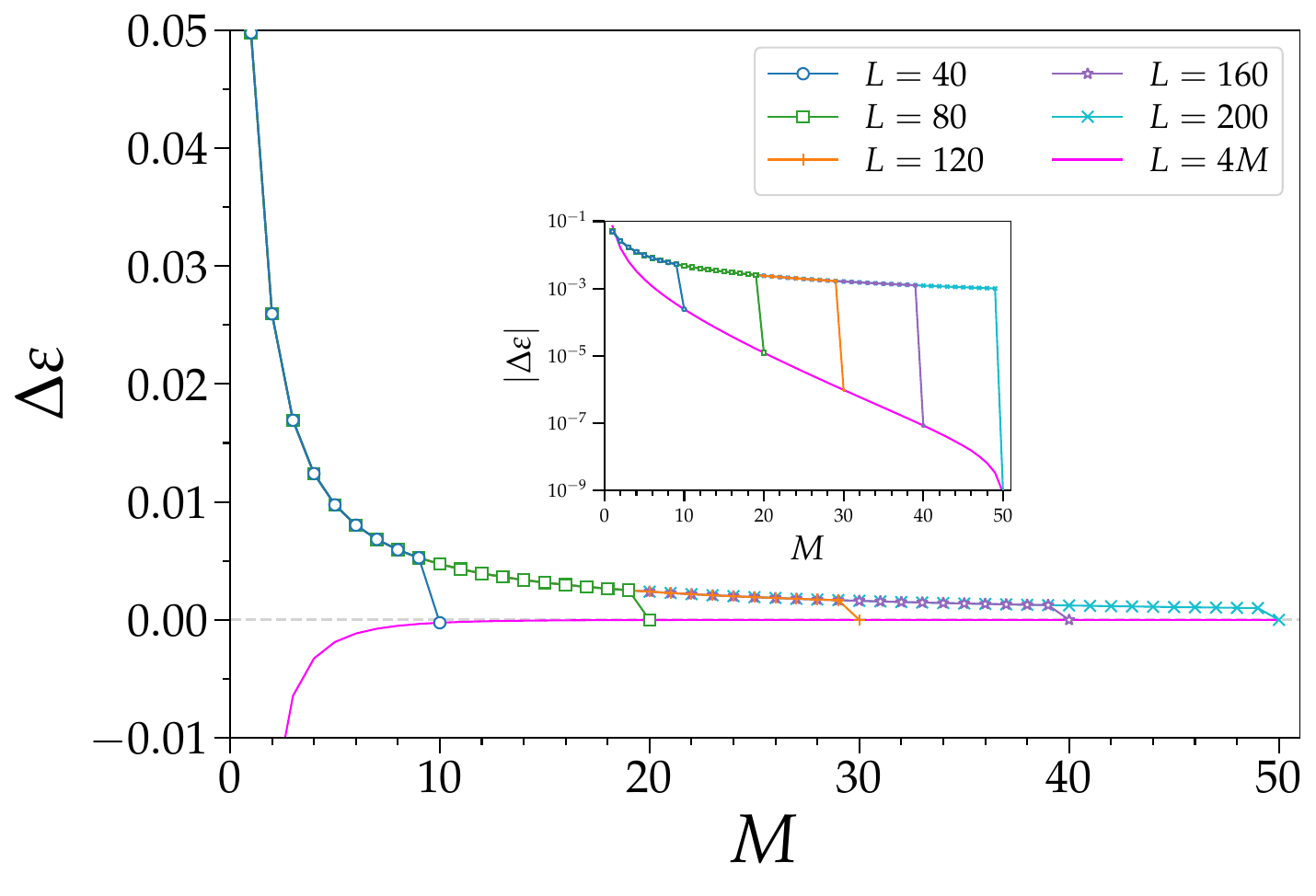}
\caption{The energy difference per site,
$\Delta \varepsilon =E_L^M(\bm\theta_{\rm opt})/L-\varepsilon_{\infty}$, for various
system sizes $L$, where $\varepsilon_{\infty}=\lim_{L\to\infty}E_{\rm exact}(L)/L$ and
$E_{\rm exact}(L)$ is the exact ground-state energy for a system size $L$. The parameters
in the final Hamiltonian are set to $(v, w) = (1, 1.1)$, and $L$ is chosen as
$L\in4\mathbb{N}$ under APBCs. The magenta line represents
$\Delta \varepsilon_A =E_{\rm exact}(L)/L-\varepsilon_{\infty}$ with $L=4M$. The inset
shows a semi-logarithmic plot.}
\label{fig:fig11}
\end{figure}

For the 1D SSH model of spinless fermions at half filling, the ground-state energy per
site $\varepsilon_{\infty}$ in the thermodynamic limit is given by
\begin{equation}
\varepsilon_{\infty} = \frac{(v+w)}{\pi} \mathcal{E}_2\left(\frac{\pi}{2}, \frac{4vw}{(v + w)^2}\right),
\end{equation}
where
\begin{equation}
\mathcal{E}_2(\phi,m) = \int_{0}^{\phi} \sqrt{1-m\sin^2\theta}d\theta
\end{equation}
denotes the incomplete elliptic integral of the second kind. Figure~\ref{fig:fig11} shows
the energy difference per site,
$\Delta\varepsilon=E_L^M(\bm\theta_{\rm opt})/L-\varepsilon_{\infty}$, between the
variational ground state and the exact ground state in the thermodynamic limit. Here,
$E_L^M(\bm\theta_{\rm opt})$ is the variational energy obtained using the DQAP ansatz with
the optimized variational parameters
$\bm\theta_{\rm opt}=\{\theta_m^{(1)}, \theta_m^{(2)}\}_{m=1}^M$, consisting of $M$
layers, for a system size $L$. We find that when $M < L/4$, $\Delta \varepsilon$ remains
independent of the system size $L$. This behavior arises because quantum entanglement in a
quantum circuit composed of local two-qubit gates propagates within a causality-cone-like
structure during time evolution. As a result, the quantum gates that contribute to the
energy expectation value are confined within this causality cone, making
$\Delta\varepsilon$ unaffected by the system size. A detailed analysis of this effect can
be found in Refs.~\cite{PhysRevResearch.3.013004} and~\cite{2019arXiv191112259B}.

\subsection{Mutual information for a DQAP ansatz with a fixed $M$} \label{app:mi2}

\begin{figure}
\includegraphics[width=0.95\columnwidth]{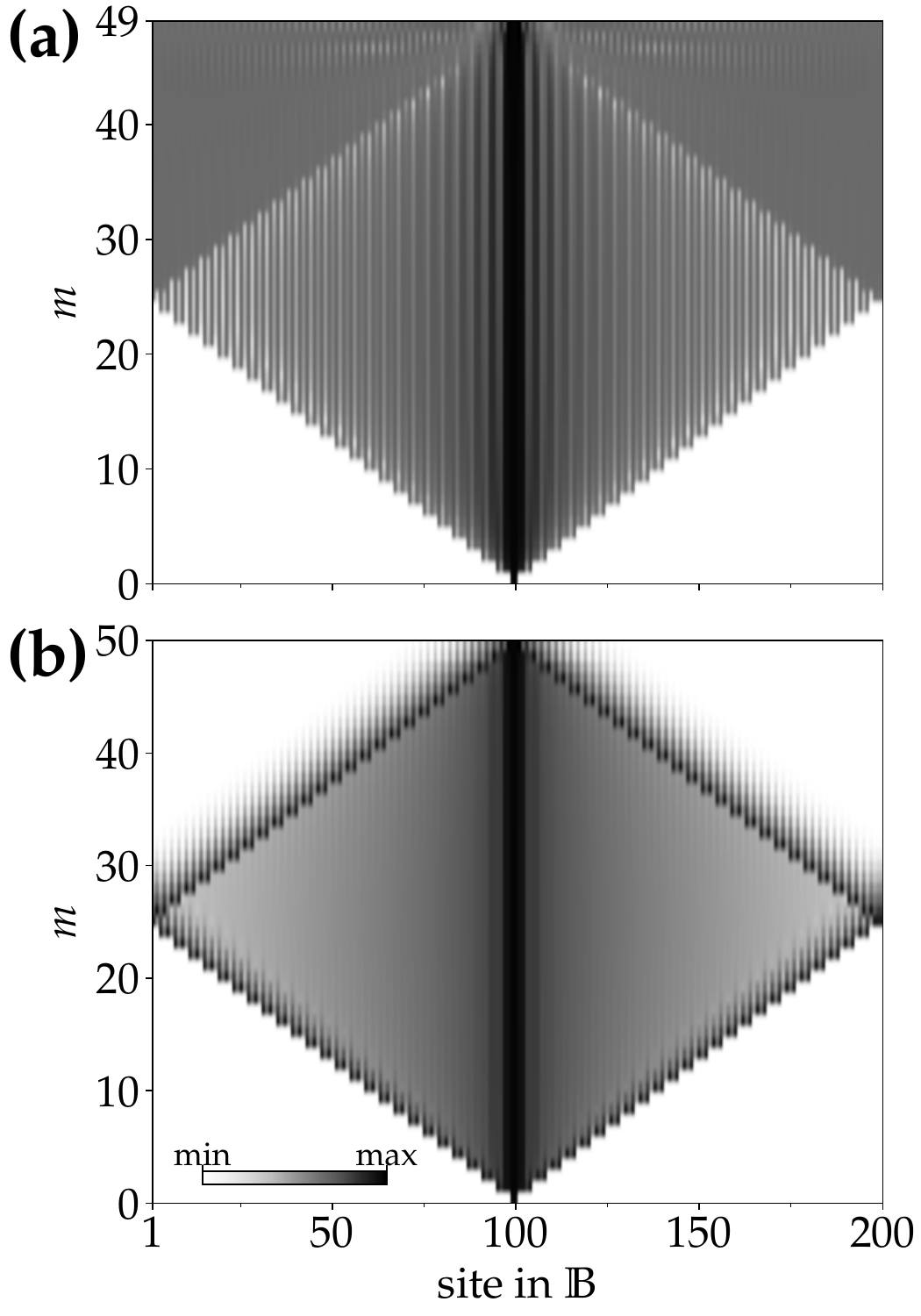}
\caption{Intensity plot of the mutual information $I_{\mathbb{A}, \mathbb{B}}$ for the
optimized DQAP ansatz $|\Psi_M(\bm\theta)\rangle$ with (a) $M=49$ and (b) $M=50$, shown as
a function of the circuit depth $m$ and the location of the single site composing
subsystem $\mathbb B$. Here, the system size is set to $L=200$ under APBCs and subsystem
$\mathbb A$ consists of two sites located at the center of the system, i.e., at the 99th
and 100th sites (see Sec.~\ref{sec:mi} and Fig.~\ref{fig:fig6}). The parameters of the
final target Hamiltonian are set to $(v, w)=(1, 2)$, corresponding to Path II in
Fig.~\ref{fig:fig1}(b). For each circuit depth $m$, only the first $m$ values of the
optimized variational parameters $\{\theta_{m'}^{(1)}, \theta_{m'}^{(2)}\}_{m'=1}^M$ are
used to compute the mutual information. In contrast, in Fig.~\ref{fig:fig6}, the
variational parameters are optimized for each $M$. White color in the intensity plots
represent zero.}
\label{fig:fig12}
\end{figure}

It is also insightful to examine how the mutual information $I_{\mathbb{A}, \mathbb{B}}$
evolves as the number $m\, (\leqslant M)$ of layers increases in the DQAP ansatz
$|\Psi_M(\bm\theta)\rangle$, where the variational parameters
$\bm\theta=\{\theta_m^{(1)}, \theta_m^{(2)}\}_{m=1}^M$ are optimized for a given $M$.
Figure~\ref{fig:fig12} shows the representative results for the mutual information
$I_{\mathbb{A}, \mathbb{B}}$ in a system of size $L=200$, where subsystem $\mathbb A$
consists of two sites located at the center (i.e., at the 99th and 100th sites), while
subsystem $\mathbb B$ contains a single site whose position is varied (see
Sec.~\ref{sec:mi} and Fig.~\ref{fig:fig6}). Note that the variational parameters in the
DQAP ansatz $|\Psi_M(\bm\theta)\rangle$ are optimized for $M=49$ in
Fig.~\ref{fig:fig12}(a), which is one layer short of reaching the exact ground state, and
for $M=50$ in Fig.~\ref{fig:fig12}(b), corresponding to the exact ground state of the
final target Hamiltonian with $(v, w)=1, 2$, which follows Path II in
Fig.~\ref{fig:fig1}(b). As shown in Fig.~\ref{fig:fig12}, in both cases, the entanglement
gradually expands in space as the number $m$ of layers increases, forming a causal cone
that defines the maximum propagation speed of information with local two-qubit gates,
until it spans the entire system at $m=L/8$. However, beyond this point, the behavior of
entanglement evolution differs: for $M=49$, the entanglement remains extended across the
whole system, while for $M=50$, as the number $m$ of layers further increases, the
entanglement gradually contracts, becoming more localized around the center of the system.

\subsection{Optimized variational parameters} \label{app:opt}

\begin{figure*}
\includegraphics[width=1.95\columnwidth]{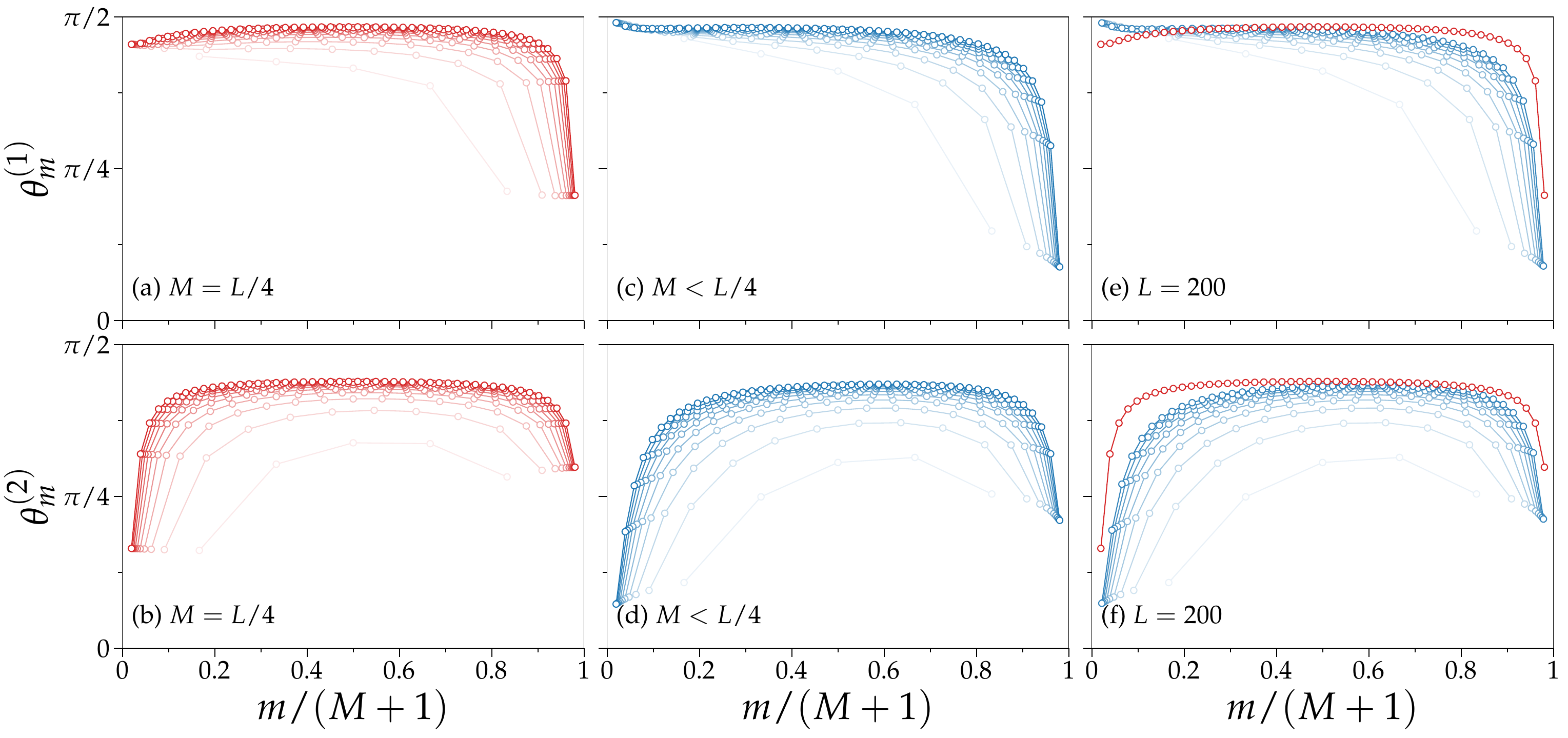}
\caption{Optimized variational parameters ${\bm\theta}_{\text{opt}}$ for
$(v, w) = (1.0, 1.1)$, corresponding to Path II. (a, b) ${\bm\theta}_{\text{opt}}$ for
various system sizes up to $L=200$ with $M = L/4$,
where the exact ground state state is successfully prepared. (c, d)
${\bm\theta}_{\text{opt}}$ for $M < L/4$ with $M$ up to 50. Note
that as long as $M<L/4$, the optimized variational parameters are independent of the
system size $L$, for which the DQAP ansatz $|\Psi_M(\bm\theta)\rangle$ with
$\bm\theta=\{\theta_m^{(1)}, \theta_m^{(2)}\}_{m=1}^M$ is optimized. (e, f)
${\bm\theta}_{\text{opt}}$ for a fixed system size $L = 200$, where the DQAP ansatz
$|\Psi_M(\bm\theta)\rangle$ is optimized for different $M$. 
A clear discontinuity in the sets of the
optimized parameters is observed between $M=50$ (indicated in red) and $M<50$ (indicated
in blue), which is consistent with the abrupt variations in the ground-state energy
[Fig.~\ref{fig:fig3}(b)], entanglement entropy [Fig.~\ref{fig:fig4}(b)], and mutual
information [Fig.~\ref{fig:fig6}(b)].
}
\label{fig:fig14}
\end{figure*}

\begin{figure*}
\includegraphics[width=1.95\columnwidth]{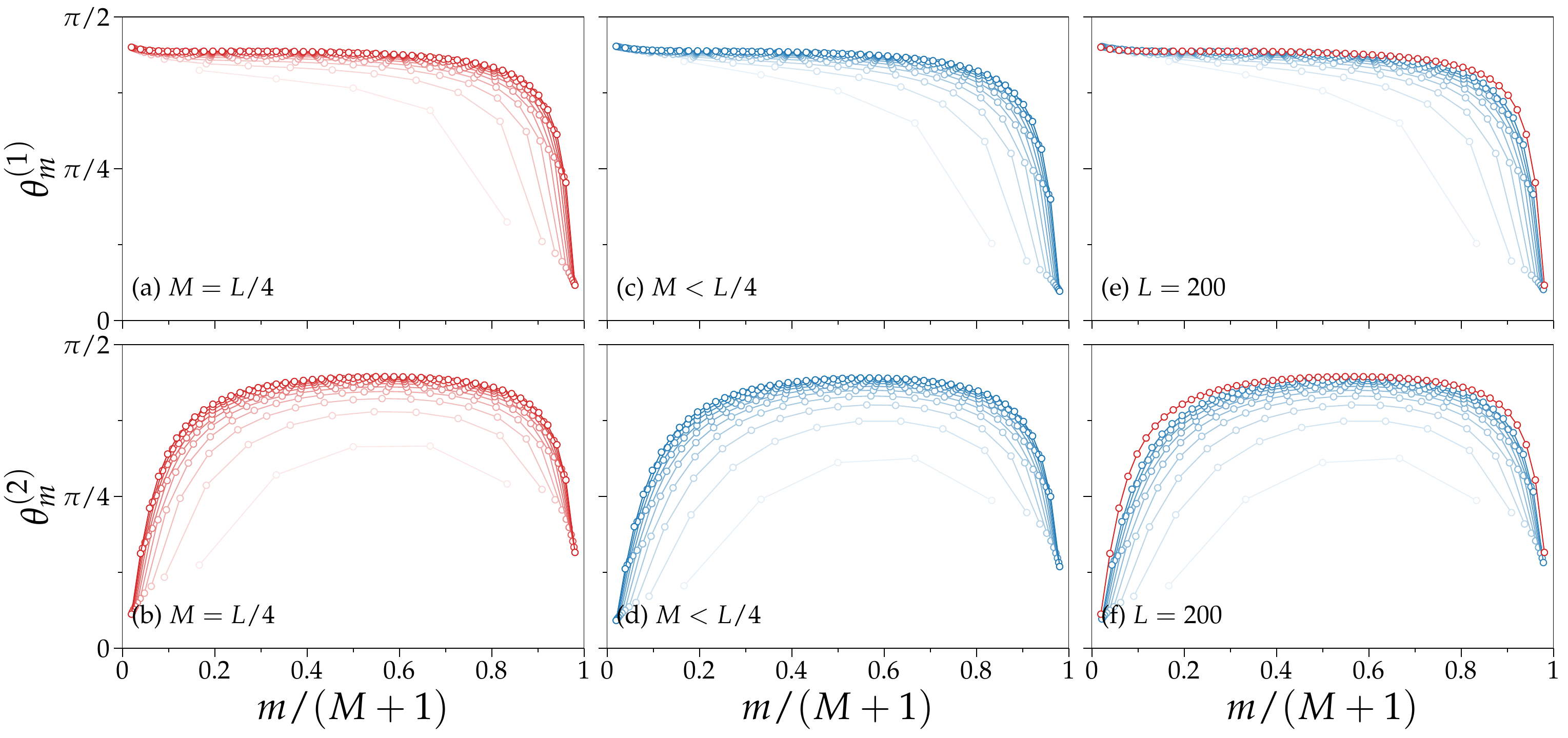}
\caption{Same as Fig.~\ref{fig:fig14}, except that the parameters in the target final
Hamiltonian are set to $(v, w) = (1.1, 1)$, corresponding to Path I.
}
\label{fig:fig15}
\end{figure*}

Figures~\ref{fig:fig14} and~\ref{fig:fig15} summarize the optimized parameters
$\bm{\theta}_{\text{opt}}$ in the DQAP ansatz $|\Psi_M(\bm\theta)$ with
$\bm\theta=\{\theta_m^{(1)}, \theta_m^{(2)}\}_{m=1}^M$ for the target final Hamiltonian
with $(v, w)=(1, 1.1)$ and $(1.1, 1)$, respectively. These correspond to Path II (where
the initial and final states belong to different topological phases) and Path I (where
both the initial and final states belong to the same topologically trivial phase) in
Fig.~\ref{fig:fig1}(b). Figures~\ref{fig:fig14}(a) and~\ref{fig:fig14}(b)
[Figures~\ref{fig:fig15}(a) and~\ref{fig:fig15}(b)] show the optimized parameters for the
case of Path II (Path I), where the system size $L\, (\in4\mathbb N)$ is varied with
$M=L/4$. Therefore, the optimized DQAP ansatz $|\Psi_M(\bm\theta)\rangle$ represents the
exact ground state of the target final Hamiltonian. We observe that these optimized
parameters vary smoothly as $L$ increases.

Figures~\ref{fig:fig14}(c) and~\ref{fig:fig14}(d) [Figures~\ref{fig:fig15}(c)
and~\ref{fig:fig15}(d)] show the optimized parameters for the case of Path II (Path I)
with $M<L/4$, where the optimized parameters in the DQAP ansatz
$|\Psi_M(\bm\theta)\rangle$ for each $M$ are independent of the system size $L$.
Specifically, the optimized parameters $\{\theta_m^{(1)}, \theta_m^{(2)}\}_{m=1}^M$ for a
system size $L_1$ are exactly the same as those for a system size $L_2$, as long as
$4M<L_1, L_2$, assuming that $L_1, L_2\in4\mathbb N$ under APBCs. This behavior arises
because the variational parameters $\bm\theta$ in the DQAP ansatz
$|\Psi_M(\bm\theta)\rangle$ are optimized to minimize the expectation value of energy for
the target final Hamiltonian, and the causality cone relevant to this energy expectation
value does not extend across the entire system as long as
$M<L/4$~\cite{PhysRevResearch.3.013004}. As shown in these figures, we also observe that
the optimized parameters vary smoothly with increasing $M$, which are clearly different
from those in Figures~\ref{fig:fig14}(a) and~\ref{fig:fig14}(b)
[Figures~\ref{fig:fig15}(a) and~\ref{fig:fig15}(b)]. Consequently, when the system size
$L$ is fixed and $M$ is varied, discontinuous changes appear in the optimized parameters
at $M=L/4$ and $M=L/4-1$, as indicated in read and blue in Figs.~\ref{fig:fig14}(e)
and~\ref{fig:fig14}(f) for the case of Path II and in Figs.~\ref{fig:fig15}(e)
and~\ref{fig:fig15}(f) for the case of Path I, although these discontinuities are less
pronounced in the latter. This discontinuity is consistent with the abrupt changes
observed in the ground-state energy [Fig.~\ref{fig:fig3}(b)], entanglement entropy
[Fig.~\ref{fig:fig4}(b)], and mutual information [Fig.~\ref{fig:fig6}(b)].

\section{Number of native two-qubit gates} \label{app:native2Qgates}

In this Appendix, we count the number of native two-qubit gates in the H1-1 system used
for the quantum circuit shown in Fig.~\ref{fig:fig9}, assuming that $L$ is even.
The number of two-qubit gates is more significant than the number of single-qubit gates,
as their infidelity is currently about two orders of magnitude larger than that of
single-qubit gates. Thus, optimizing two-qubit gate usage is crucial for improving overall
circuit fidelity.

The native two-qubit gate of the H1-1 system is the ZZPhase gate,
${\rm ZZPhase}(\alpha) = e^{-\frac{1}{2} i \pi \alpha \hat{Z}_i \hat{Z}_j}$. For
convenience, we introduce the ISWAP gate,
${\rm ISWAP(\alpha)}= e^{-\frac{1}{4} i \pi \alpha (\hat{X}_i \hat{X}_j+\hat{Y}_i \hat{Y}_j)}$,
as defined in TKET~\cite{Sivarajah_2020}. When it is compiled for the H1-1 system, a
single ISWAP gate is decomposed into two ZZPhase gates, supplemented with appropriate
single-qubit rotation gates.


%
First, $L/2$ ZZPhase gates are required to implement the initial state
$|\Psi^{(1)}\rangle$ in Eq.~(\ref{eq:DQAP}), since a single CNOT gate is equivalent to a
single ${\rm ZZPhase}(1/2)$ gate, up to single-qubit rotations.
Second, $2ML$ ZZPhase gates are needed to implement the $M$-layer DQAP ansatz unitary.
This is because the DQAP ansatz unitary consists of $ML$ ISWAP gates arranged in a
brick-wall manner (see Fig.~\ref{fig:fig9}).
Third, $L-1$ ZZPhase gates are required to implement the unitary
${\rm C}_0{\text -}\hat{\cal U}_{\rm R}$ in Eq.~(\ref{eq:C-U}), as a single
controlled-phase gate is equivalent to a single ZZPhase gate, up to single-qubit
rotations.
Thus, the total number of ZZPhase gates in the Hadamard-test circuit
(Fig.~\ref{fig:fig9}) is given by
\begin{equation}
N_{\rm ZZPhase} = 2ML + \frac{3L}{2}-1.
\end{equation}
For a fixed system size of $L=18$, we find that $N_{\rm ZZPhase}=$
26, 62, 98, 134, and 170 for $M=$ 0, 1, 2, 3, and 4, respectively.
Importantly, no SWAP gates are used in the circuit due to the all-to-all connectivity of
the H1-1 system.

\section{Additional experimental results}\label{app:experiments}

In this Appendix, we present additional experimental results for the 
energy and polarization of the $L=18$ system, obtained using 
Quantinuum's trapped-ion quantum computer \textit{Reimei}.

The experiments were conducted in March 2025. 
At the time of the experiments, the \textit{Reimei} system consisted 
of $20$ qubits and natively supported single-qubit rotation gates and 
two-qubit ZZ phase gates, defined as ${\rm ZZPhase}(\alpha) = e^{-\frac{1}{2} i \pi \alpha \hat{Z}_i \hat{Z}_j}$, where $\alpha$ 
is a real-valued parameter. These native two-qubit gates 
could be applied to arbitrary pairs of qubits. 
The average infidelity of single-qubit and two-qubit gates was 
approximately $0.007\%$ and $0.14\%$, respectively, while the average 
state preparation and measurement (SPAM) error was around $0.35\%$. 
Further details on the hardware specifications can be found in 
Ref.~\cite{reimei_spec}. 
All quantum circuits used in the experiments were compiled using 
TKET~\cite{Sivarajah_2020}.

\begin{figure}
\includegraphics[width=0.95\columnwidth]{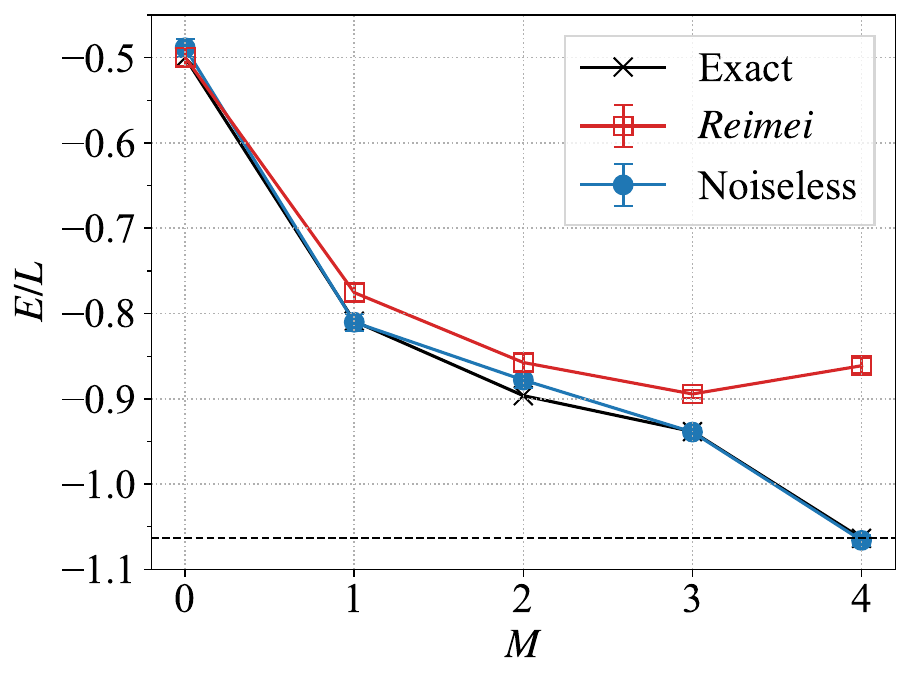}
\caption{Energy per site as a function of the circuit depth $M$ 
for the 18-site lattice. 
Red squares represent the experimental results obtained using the 
\textit{Reimei} system, 
blue circles denote the noiseless simulation results, and 
black crosses indicate the numerically exact results. 
The dashed horizontal line represents the exact ground-state energy 
of the final target Hamiltonian. 
The model parameters are the same as those used 
in Fig.~\ref{fig:figReIm}.}
\label{fig:fig13}
\end{figure}

Figure~\ref{fig:fig13} shows the energy per site evaluated using 
the \textit{Reimei} system. 
The model parameters are the same as those used in the 
Fig.~\ref{fig:figReIm}. 
Under the JWT, the hopping term between sites $l$ and $l+1$ is 
expressed as
\begin{equation}
    \hat{c}_l^\dag \hat{c}_{l+1} + {\rm H.c.} = \frac{1}{2}
    \left(\hat{X}_l \hat{X}_{l+1} + \hat{Y}_l \hat{Y}_{l+1}\right).
\end{equation}
To evaluate the expectation values of $\hat{X}_l \hat{X}_{l+1}$ and 
$\hat{Y}_l \hat{Y}_{l+1}$ for $l=1,2,\cdots,L$, we perform 
measurements of all qubits in the $X$ and $Y$ basis, respectively, 
with respect to the optimized DQAP ansatz state 
$|\Psi_{M}(\boldsymbol{\theta})\rangle$.
As in Appendix~\ref{app:native2Qgates}, 
the number of ZZPhase gates required to prepare the state 
$|\Psi_{M} \boldsymbol{\theta})\rangle$ is given by 
\begin{equation}
N_{\rm ZZPhase}^\prime = 2ML + \frac{L}{2},
\end{equation}
where the first term accounts for the number of ZZPhase gates in the 
$M$-layer DQAP ansatz unitary, and the second term accounts for 
those required for preparing the initial state $|\Psi^{(1)}\rangle$. 
The difference $N_{\rm ZZPhase}-N_{\rm ZZPhase}^\prime=L-1$ 
corresponds to the number of controlled-phase gates used in the 
Hadamard-test circuit for polarization calculations. 
For the system size $L=18$, we find that $N_{\rm ZZPhase}^\prime=$ 
9, 45, 81, 117, and 153 for $M=$ 0, 1, 2, 3, and 4, respectively.
The energy per site is then evaluated accordingly to 
Eqs.~(\ref{eq:H_SSH_JWT}) and (\ref{eq:energy}).
We observe that the energy is significantly larger than the exact 
value for $M=1,2,3$ and $4$. 
Moreover, while the energy decreases with increasing $M$ up to $M=3$, 
it increases at $M=4$. 
In particular, the energy at $M=4$ is noticeably lager than the exact 
value and the result from noiseless simulation.

\begin{figure}
\includegraphics[width=0.95\columnwidth]{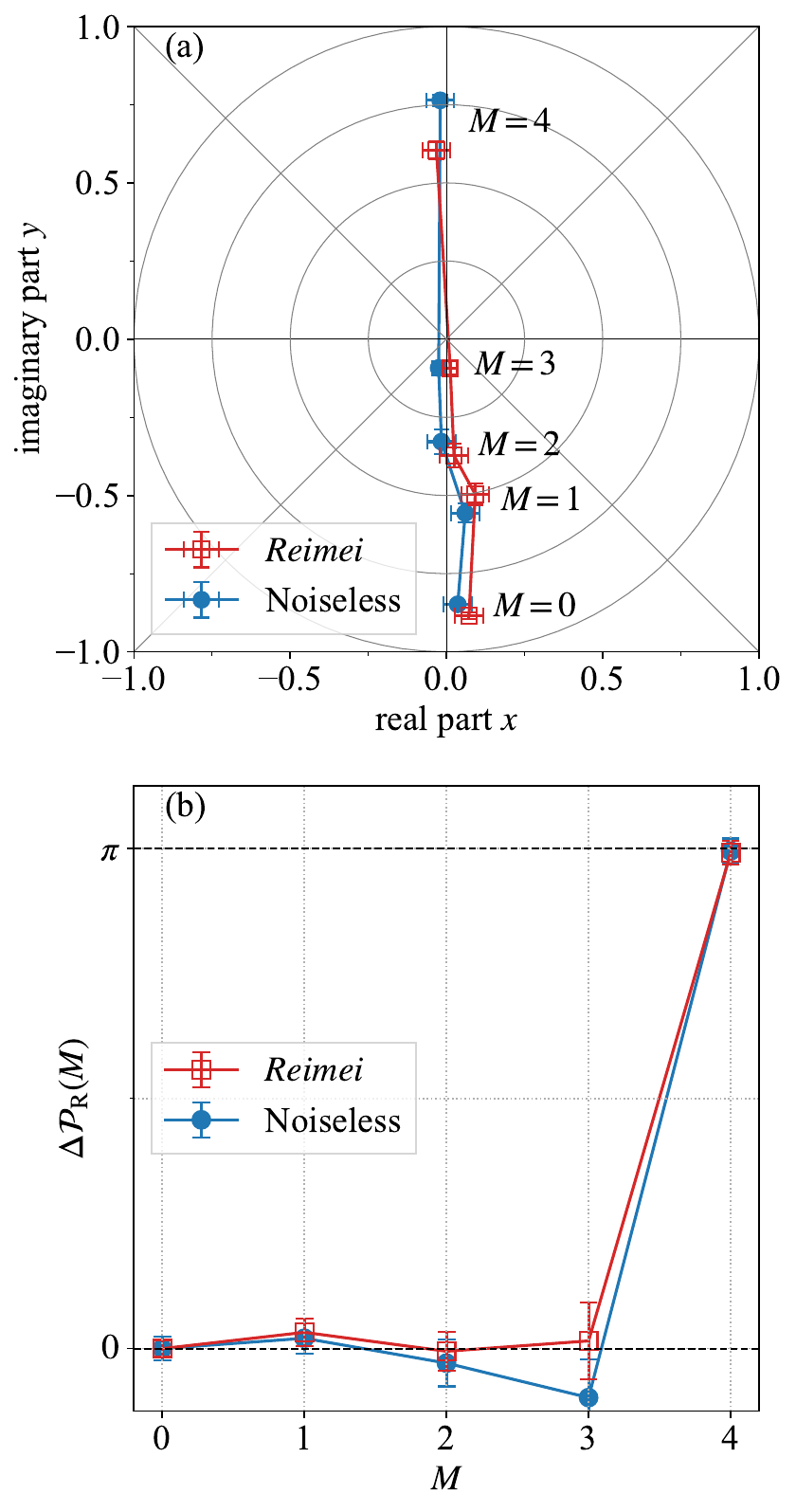}
\caption{
Same as Fig.~\ref{fig:figReIm}, but the experimental results are 
obtained using the \textit{Reimei} system.
}
\label{fig:figReIm_reimei}
\end{figure}

Figure~\ref{fig:figReIm_reimei} shows the real and imaginary parts of $\langle \Psi_{M}(\boldsymbol{\theta}_{\rm opt})|\hat{\cal U}_{\rm R} |\Psi_{M}(\boldsymbol{\theta}_{\rm opt})\rangle$ as well as the 
associated polarization difference $\Delta {\cal P}_{\rm R}(M)$, 
evaluated in the same way as in Fig.~\ref{fig:figReIm} but using 
the \textit{Reimei} system. 
Despite the significantly larger energy values compared to the exact 
results for $M=1,2,3,$ and $4$, the polarization evaluated with the 
\textit{Reimei} system successfully captures the topological phase 
transition. 
These results support the robustness of the polarization against 
noise, attributed to its topological nature, as discussed in 
Sec.~\ref{sec:qc}. 
It should be noted, however, that the energy and polarization 
experiments shown in Figs.~\ref{fig:fig13} and 
\ref{fig:figReIm_reimei}, respectively, were performed independently.

\bibliography{refs}


\end{document}